\documentclass[11pt,a4paper]{article}

\pdfoutput=1
\usepackage{jheppub}
\usepackage{graphicx,color,float}
\usepackage{hyperref}
\usepackage{multirow}
\usepackage{verbatim}
\usepackage{longtable}
\usepackage{amsmath}
\usepackage{amsfonts}
\usepackage{amssymb}
\usepackage{rotating}

\usepackage[normalem]{ulem}

\usepackage{siunitx}

\def\gsim{\raise0.3ex\hbox{$\;>$\kern-0.75em\raise-1.1ex\hbox{$\sim\;$}}}
\def\lsim{\raise0.3ex\hbox{$\;<$\kern-0.75em\raise-1.1ex\hbox{$\sim\;$}}}

\begin{document}


\title{Probing dark photons from a light scalar at Belle II}

\author[a,b,c]{Kingman Cheung,}
\emailAdd{cheung@phys.nthu.edu.tw}
\affiliation[a]{Department of Physics, National Tsing Hua University,	Hsinchu 300, Taiwan}
\affiliation[b]{Center for Theory and Computation, National Tsing Hua University,	Hsinchu 300, Taiwan}
\affiliation[c]{Division of Quantum Phases and Devices, School of Physics, Konkuk University, Seoul 143-701, Republic of Korea}

\author[d]{Yongkyu Kim,}
\emailAdd{ykjk1401@yonsei.ac.kr}
\affiliation[d]{Department of Physics, Yonsei University, Seoul 03722, Republic of Korea}

\author[d]{Youngjoon Kwon,}
\emailAdd{yjkwon63@yonsei.ac.kr}

\author[a,b]{C.J.~Ouseph,}
\emailAdd{ouseph444@gmail.com}

\author[e]{Abner Soffer,}
\emailAdd{asoffer@tau.ac.il}
\affiliation[e]{School of Physics and Astronomy, Tel Aviv University, Tel Aviv 69978, Israel}

\author[a,b]{Zeren Simon Wang}
\emailAdd{wzs@mx.nthu.edu.tw}

\date{\today}

\vskip1mm
\abstract{In the minimal $U(1)$ extension of the Standard Model (SM), a new gauge boson referred to as ``dark photon'' is predicted.
The dark-photon mass can be generated from an additional Higgs mechanism associated with a dark scalar boson.
At $B$-factories such as Belle II, large numbers of $B$-mesons are produced and can decay to a kaon plus the dark scalar via the latter's mixing with the SM Higgs boson.
We evaluate the sensitivity of Belle II for the case in which the dark scalar decays exclusively into a pair of dark photons via the new $U(1)$ gauge coupling, and the dark photons are long lived owing to a small kinetic mixing $\epsilon$.
We study the experimental signature in which each dark photon decays into a pair of charged leptons, pions, or kaons, resulting in a pair of displaced vertices, and argue that the search is essentially background-free.
We perform detailed Monte-Carlo simulations to determine the expected number of signal events at Belle II with an integrated luminosity of 50 ab$^{-1}$, taking into account the efficiencies for both final-state-particle identification and displaced tracking.
We find that for experimentally allowed values of the scalar mixing angle and kinematically allowed dark-photon and dark-scalar masses, the proposed search is uniquely sensitive to the medium-$\epsilon$ regime, which is currently mostly unexcluded by experiments.
}



\maketitle

%


\section{Introduction}\label{sec:intro}

The existence of dark matter (DM) is strongly indicated by all astronomical observations, notably gravitational lensing, galactic rotation curves, the Bullet Cluster, and CMB measurements.
Nevertheless, the nature and identity of the DM are entirely unknown.
Most DM-particle searches have focused on a weakly interacting massive particle with mass $\mathcal{O}(10-1000)$~GeV (see for example Ref.~\cite{Schumann:2019eaa}).
Lack of a discovery in this scenario has expanded the interest to various DM-sector particles, defined as those that interact with the stable DM, and which may have mass of order GeV or even below.
Typical examples of such particles include axions, axion-like particles, dark gauge bosons, dark scalars, and dark fermions~\cite{Beacham:2019nyx,Agrawal:2021dbo,Antel:2023hkf}.

Dark-sector models generically posit the existence of gauge bosons, scalar bosons, and fermions with a symmetry under which the Standard-Model (SM) particles are singlets.
One of the simplest examples is a new gauge boson associated with a new $U(1)$ symmetry.
Such a dark gauge boson, coined ``dark photon''~\cite{Curtin:2014cca}, can mix with the SM photon via a kinetic mixing term, $\epsilon F^{\mu\nu} F'_{\mu\nu}$, where $F_{\mu\nu} $ and $F'_{\mu\nu}$ denote the field strength of the SM photon and dark photon fields, respectively, and $\epsilon$ is the mixing coefficient.
Kinetic mixing enables the dark photon mass eigenstate $\gamma'$ to interact with SM particles, facilitating its creation and detection in experiments.

Various terrestrial experiments have established constraints on a dark photon with a mass $\gtrsim 1$~MeV. 
Signatures in which the dark-photon decay vertex is prompt or slightly displaced with respect to the beams interaction point (IP) were utilized in collider searches~\cite{CMS:2018rdr,CMS:2018lqx,ATLAS:2015itk,ATLAS:2014fzk,ATLAS:2016jza,LHCb:2017trq,BaBar:2014zli}.
Signatures with highly displaced decays were studied at fixed-target and beam-dump experiments~\cite{Riordan:1987aw,Bross:1989mp,Davier:1989wz,NA482:2015wmo}.
In addition, bounds on energy losses in supernovae impose further limits in the region of small masses $m_{\gamma '}\lesssim \mathcal{O}(10^{-1})$ GeV.
These limits were discussed in Refs.~\cite{Dent:2012mx,Dreiner:2013mua} and updated in Refs.~\cite{Chang:2016ntp,Hardy:2016kme,Bottaro:2023gep} to include the effect of finite temperature and plasma density as well as white dwarf bremsstrahlung.
Also, the electron magnetic moment, with its very precise experimental determination, has been used to set an indirect limit~\cite{Pospelov:2008zw}.
For the mass range $m_{\gamma'}\sim 1~\rm MeV - 100~\rm MeV$, limits around $\epsilon \lesssim 10^{-10}$ have been determined from cosmology, arising from the cosmic microwave background and nucleosynthesis~\cite{Fradette:2014sza}.
For a comprehensive list of the past experimental searches, we refer the reader to Refs.~\cite{Bauer:2018onh,Caputo:2021eaa}\footnote{See also \url{https://cajohare.github.io/AxionLimits/docs/dp.html} for a graphical compilation of these existing limits.} and the references therein.
Ref.~\cite{Ilten:2018crw} provides a summary of both current and prospective constraints on dark photons, as well as software for re-evaluating various dark-photon models.

In addition, a ``dark scalar'' or ``dark-Higgs boson'', denoted $\phi_D$, is naturally present in dark-sector models, particularly for generating the mass of the dark photon.
The dark Higgs can couple to the SM Higgs doublet via a renormalizable term $\lambda(H^\dagger H)(\phi^\dagger_D \phi_D)$. 
These scalars are related through a mixing angle $\theta$ to the mass eigenstates $\phi$ and $h_{125}$, the latter being the scalar boson observed at the LHC.
For the relation between $\lambda$ and $\theta$, see, for instance, Eq.~(9) in Ref.~\cite{Cheung:2019qdr}.
The scalar mixing angle $\theta$ is constrained to be small by the LHC Higgs boson data~\cite{Cheung:2015dta}, a number of fixed-target \cite{osti_1440463,CHARM:1985anb,NA62:2020pwi,NA62:2021zjw,BNL-E949:2009dza,Winkler:2018qyg,KTEV:2000ngj,Dolan:2014ska,Gorbunov:2021ccu,KOTO:2020prk,Egana-Ugrinovic:2019wzj,MicroBooNE:2021usw} and collider~\cite{LHCb:2015nkv, LHCb:2016awg,Belle-II:2023ueh} experiments, as well as astrophysical observations of the supernovae SN1987a~\cite{Raffelt:1996wa,Krnjaic:2015mbs,Kamiokande-II:1987idp,Dev:2020eam} and cooling in stars~\cite{Hardy:2016kme}.
For a summary of these existing constraints, see Ref.~\cite{Ferber:2023iso}.

Theoretical scenarios involving both the dark Higgs and the dark photon have been extensively studied from the collider-phenomenological and cosmological perspectives~\cite{Wells:2008xg, Ahlers:2008qc,Gopalakrishna:2008dv, Batell:2009yf,Weihs:2011wp, Davoudiasl:2013aya,Curtin:2013fra,Falkowski:2014ffa,Bakhet:2015pqa,Izaguirre:2018atq,Curtin:2014cca,Jodlowski:2019ycu,Araki:2020wkq,Foguel:2022unm}, and signals have also been searched for at $B$ factories~\cite{BaBar:2012bkw,Jaegle:2015fme,Belle:2020the,Belle-II:2022jyy}.
Since the $B$-factory searches considered a different theoretical model from the one studied here, we find that their published bounds, both model-independent and model-dependent, cannot be reinterpreted as constraints on the model considered here.
Therefore, we do not include them in the numerical results.

We note that the couplings of the dark Higgs\footnote{For clarification, we use the terms ``dark Higgs/scalar'' and ``dark photon'' to denote both the weak eigenstate and the mass eigenstate interchangeably, of the corresponding fields.} to the SM fermions and to the $W$- and $Z$-bosons are governed by the mixing angle $\theta$, while the couplings of the dark photon to the SM fermions are dictated by $\epsilon$. 
In addition, importantly, the dark Higgs couples to a pair of dark photons via the new gauge coupling associated with the appended $U(1)$ symmetry.
This facilitates the decoupling of the production rate and the lifetime of the dark photon when it is produced in dark-scalar decays.
In this paper we focus on the case in which $\phi\to \gamma'\gamma'$ is by far the dominant decay mode of the dark scalar, and the dark Higgs decays promptly.

Sensitive searches for a dark scalar and dark photon with masses around a GeV can be carried out at $B$-factories. 
For example, the Belle~II experiment~\cite{Belle-II:2010dht,Belle-II:2018jsg} plans to produce as many as $5.5\times 10^{10}$ $B$-meson pairs in the coming decade.
As a result, it can search for rare decays of the $B$-meson with branching ratios as small as $\mathcal{O}(10^{-10})$, as long as the search channel has reasonably high efficiency and very low background. 

In this work, we study the decay chain:
\[
  B^\pm \to K^\pm \phi, \, \, \, \, \, \phi \to  \gamma' \gamma',
\]
followed by the displaced decay of each dark photon 
\[
  \gamma' \to e^+ e^-,\, \mu^+ \mu^-,\, \pi^+ \pi^-,\, K^+ K^- \;.
\]
This signal process was first proposed in Ref.~\cite{Batell:2009jf} for study at $B$-factories (see also e.g.~Refs.~\cite{Ferber:2022ewf,Bandyopadhyay:2023lvo,Bandyopadhyay:2022klg,Batell:2022dpx,Jaeckel:2023huy} for some recent phenomenological studies on Belle~II sensitivities to long-lived dark photons).
However, in this work we show how to improve the sensitivity to the kinetic mixing parameter $\epsilon$ down to $\mathcal{O}(10^{-7})$ for $m_\phi=4$~GeV with $\theta = 10^{-4}$.
The sensitivity reach varies mildly for $m_\phi$ in the range $0.1 - 4$ GeV.
An important factor leading to the high sensitivity is the focus on a region of parameter space in which the dark photons are long lived and their decays produce a pair of displaced vertices (DVs) in the detector.
As we will show, our proposed search, compared to past experiments and other proposed searches, is sensitive to a unique parameter region in the medium-$\epsilon$ regime which is largely un-probed.
Further, we emphasize that this is the first study of a search for dark photons at Belle~II associated with a signature of double DVs consisting of two tracks each.

The organization of this paper is as follows.
We lay out the basics of the model in Sec.~\ref{sec:model}.
In Sec.~\ref{sec:experiment_simulation}, we introduce the Belle~II experiment, discuss the signal-event reconstruction and background sources at the experimental level, and describe the signal-event simulation procedure.
Sec.~\ref{sec:results} contains our numerical results of the Belle~II sensitivity reach in terms of the dark-photon and dark-Higgs parameters.
We summarize the work in Sec.~\ref{sec:Conclusions}. 
Additionally, in Appendix~\ref{app:efficiencies_pid}, we report the detector efficiencies that we estimate for the different final states.

\section{Model basics}\label{sec:model}

We explore an extension of the SM with an additional, dark-sector $U(1)_D$ gauge symmetry, under which all SM particles are neutral.
The gauge boson associated with this symmetry is referred to as the dark photon and denoted $\gamma'$ in this paper.
The $U(1)_D$ symmetry undergoes spontaneous breaking through the vacuum expectation value (vev) of a complex scalar field $\phi_D$, which carries a $U(1)_D$ charge.
As a result, the dark photon gains mass $m_{\gamma'}$.
Furthermore, an interaction term between two dark photons and the $CP$-even component of $\phi_D$ appears.
The interaction strength is determined by $m_{\gamma'}$  and the $U(1)_D$ gauge coupling $g'$.
Following the electroweak symmetry breaking, the dark photon can mix with the photon through the gauge kinetic mixing term between the SM hypercharge and the $U(1)_D$ gauge field.
We label the coefficient of the kinetic-mixing term with $\epsilon$.
The dark photon can thus interact with the SM fermions through the electromagnetic current.
Taking the scalar mixing angle $\theta$ to be small, the interaction Lagrangian is~\cite{Araki:2020wkq}
\begin{equation}
    \mathcal{L}_{\rm int}=g'm_{\gamma'}\phi\gamma'_\mu\gamma'^{\mu}-\epsilon e \gamma'_\mu J^\mu_{\rm EM}+\sum_f\frac{m_f\theta}{v}\phi\bar{f}f,
\end{equation}
where $v$ is the SM Higgs vev, $e=\sqrt{4\pi \alpha_{\rm QED}}$ is the electromagnetic coupling with $\alpha_{\rm QED}$ being the fine-structure constant, $J^\mu_{\text{EM}}$ denotes the SM electromagnetic current, and $f$ labels each SM fermion with mass $m_f$.

At $B$-factories, the scalar mass eigenstate $\phi$ is dominantly produced through mixing with the SM Higgs in penguin, $b\to s\phi$ decays of $B$-mesons.
We consider only the experimentally favorable $B^+\to K^+\phi$ decay (and the charge-conjugated channel). 
For the computation of this decay's width, we follow Ref.~\cite{Winkler:2018qyg}:
\begin{eqnarray}
\Gamma(B^+\to K^+\phi)&\simeq&|g_{\phi s b}|^2 \, \Big|<K^+|\bar{s_L}b_R|B^+>\Big|^2\frac{\lambda^{1/2}_{B^+,K^+\phi}}{16\pi m_{B^+}},\label{eqn:GammaB2Kphi}
\end{eqnarray}
where
\begin{eqnarray}
g_{\phi s b}&=&\frac{\theta \, m_b}{v}\frac{3\sqrt{2} \, G_F \, m^2_t \, V^*_{ts} \, V_{tb} }{16\pi^2}, \\
\lambda_{x,yz}&\equiv&\frac{m_x^2-(m_y-m_z)^2}{m_x^2}~\frac{m_x^2-(m_y+m_z)^2}{m_x^2},
\end{eqnarray}
with $m_{b/t}$ denoting the bottom/top quark mass, $G_F$ being the Fermi constant, and $V_{ts}$ and $V_{tb}$ being CKM matrix elements.
The $B^+ \to K^+$ transition matrix element can be approximated as~\cite{Ball:2004ye,Ball:2004rg}
\begin{equation}
\begin{aligned}
|<K^+|\bar{s}_Lb_R|B^+>|^2 &= \frac{1}{4}\frac{(m_{B^+}^2-m_{K^+}^2)^2}{(m_b-m_s)^2}f_K^2,
\end{aligned}
\end{equation}
with
\begin{equation}\label{eq:fK}
\begin{aligned}
    f_K &=\frac{0.33}{1-q^2/37.5~\text{GeV}^2},
\end{aligned}
\end{equation}
where $q^2 =m_\phi^2$ is the transferred momentum squared.

The partial decay widths of the dark Higgs into dark photons and charged leptons are~\cite{Araki:2020wkq}.
\begin{eqnarray}
    \Gamma(\phi\to \rm~\gamma'\gamma')&=&\frac{g'^2}{8\pi}\frac{m^2_{\rm \gamma'}}{m_\phi}\Bigg(2+\frac{m_\phi^4}{4m^4_{\gamma'}}\Big(1-\frac{2 m^2_{\rm \gamma'}}{m^2_\phi}\Big)^2\Bigg)\beta_\phi(\rm \gamma'),\label{PW_h}\\
    \Gamma(\phi\to f\bar{f})&=& N_c \,    \frac{m_\phi}{8\pi}\Big(\frac{m_f}{v}\Big)^2\theta^2\Big(1-\frac{4m_f^2}{m_\phi^2}\Big)\beta_\phi(f), \label{PW_h2}
\end{eqnarray}
where $\beta_i(j)=\sqrt{1-4m_j^2/m_i^2}$ is the kinematic factor of the decay $i\to jj$, and $N_c = 1 \,(3)$ for
$f$ being a charged lepton (quark).
It is noteworthy that the $\phi$ decay widths into SM fermions are suppressed by the scalar mixing $\theta$ (see Sec.~\ref{sec:intro}).
Thus, $\phi$ dominantly decays into a pair of dark photons~\cite{Chang:2013lfa,Araki:2020wkq}.
In practice, we simply assume that the dark Higgs boson $\phi$ decays promptly into a pair of dark photons with a branching ratio of $100\%$.
This is justified with the findings in, for instance, Ref.~\cite{Araki:2020wkq}, which shows that for $m_\phi=2$ GeV and $\theta=10^{-4}$, even setting $g'$ to be as small as $10^{-4}$ renders $\mathcal{B}(\phi\to \gamma'\gamma')\simeq 100\%$ across almost the whole kinematic range of $m_{\gamma'}$ (see also Ref.~\cite{Foguel:2022unm} for a relevant discussion).

The partial decay widths of the dark photon into a pair of charged leptons or any hadronic state $h$ (including  the two-body states of interest $\pi^+ \pi^-$ and $K^+ K^-$) are~\cite{Batell:2009yf, Bauer:2018onh,Araki:2020wkq,DOnofrio:2019dcp,Fabbrichesi:2020wbt}
\begin{eqnarray}
    \Gamma(\gamma '\to l^+l^-)&=&\frac{1}{3}~\alpha_{\rm QED}~m_{\gamma '}~\epsilon^2\sqrt{1-\frac{4m_l^2}{m^2_{\gamma '}}}\Big(1+\frac{2m_l^2}{m^2_{\gamma '}}\Big), \label{eqn:PW_A2ll}\\
    \Gamma(\gamma'\to h)&=&\Gamma(\gamma '\to \mu^+\mu^-)\times\mathcal{R}_h(s=m^2_{\gamma'}), \label{eqn:PW_A2had}
\end{eqnarray}
where the cross-section ratio $\mathcal{R}_h(s)=\sigma_{e^+e^-\to h}/\sigma_{e^+e^-\to\mu^+\mu^-}$ is extracted from Ref.~\cite{Ilten:2018crw}.
The total width of the dark photon, $\Gamma_{\gamma'}$, is the sum of Eq.~\eqref{eqn:PW_A2ll} and Eq.~\eqref{eqn:PW_A2had}, with $h$ referring to all kinematically allowed hadronic states.

\begin{figure}[t]
	\centering
	\includegraphics[width=\textwidth]{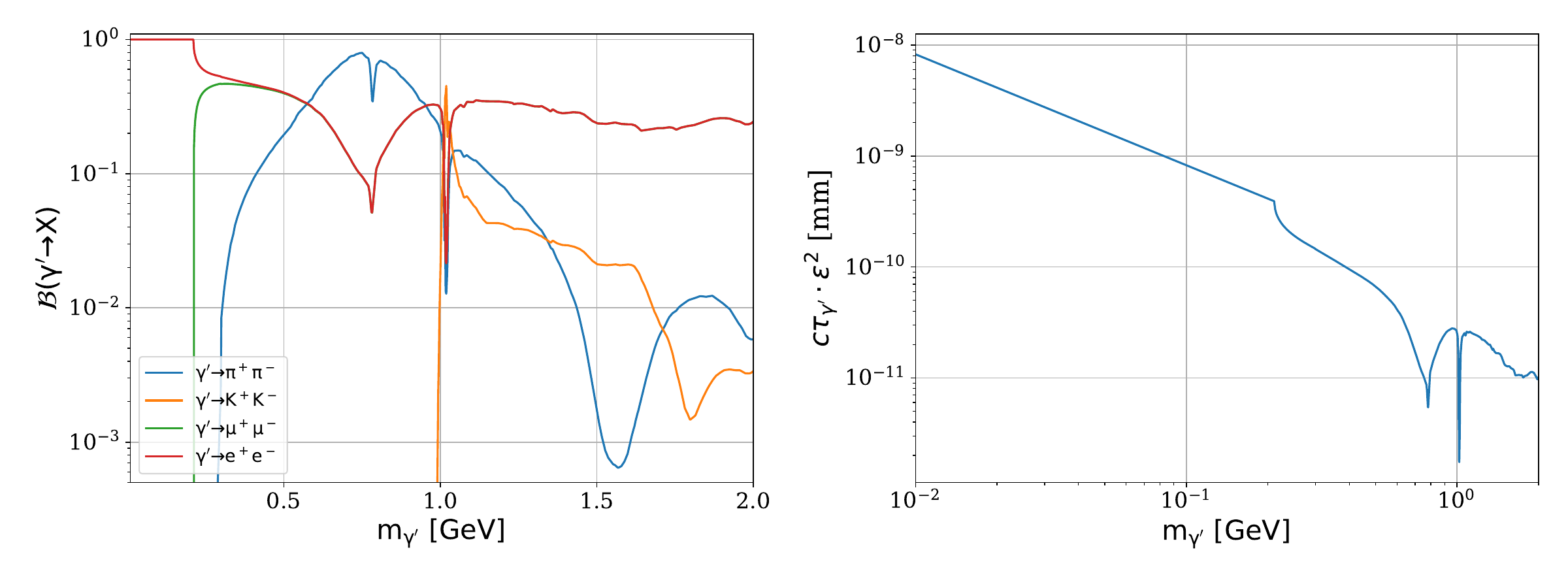}
	\caption{The dark-photon decay branching ratios into the signature final states (left), and its proper decay length re-scaled with the squared mixing coefficient $\epsilon^2$ (right), as functions of the dark-photon mass.  
 Note that in the left panel, the curves for electrons and muons overlap for $m_{\gamma'} \gtrsim 0.5$~GeV. 
The dip around 0.78 GeV is due to $\rho$ and $\omega$ mesons.
Wiggles visible mostly for $m_{\gamma'} > 1$ GeV arise from experimental fluctuations in the measurement of ${\cal R}_h$~\cite{ParticleDataGroup:2020ssz}.
For discussion of further features on these curves, we refer to Ref.~\cite{Ilten:2018crw}.
}
\label{fig:br}
\end{figure}

In the left panel of Fig.~\ref{fig:br} we show the resulting branching fractions into the signature final states, $e^+ e^-, \mu^+\mu^-, \pi^+\pi^-$, and $K^+K^-$, as functions of the dark-photon mass $m_{\gamma'}$.
In addition, in the right panel of Fig.~\ref{fig:br}, we plot the $m_{\gamma'}$ dependence of $c\tau_{\gamma'}\cdot \epsilon^2$, the proper decay length of the dark photon normalized with respect to the squared mixing parameter. 
Here, $c\tau_{\gamma'}=\hbar c/\Gamma_{\gamma'}$, where $c$ and $\hbar$ label the speed of light and the reduced Planck constant, respectively.

\begin{figure}[t]
\centering
 \includegraphics[width=0.495\textwidth]{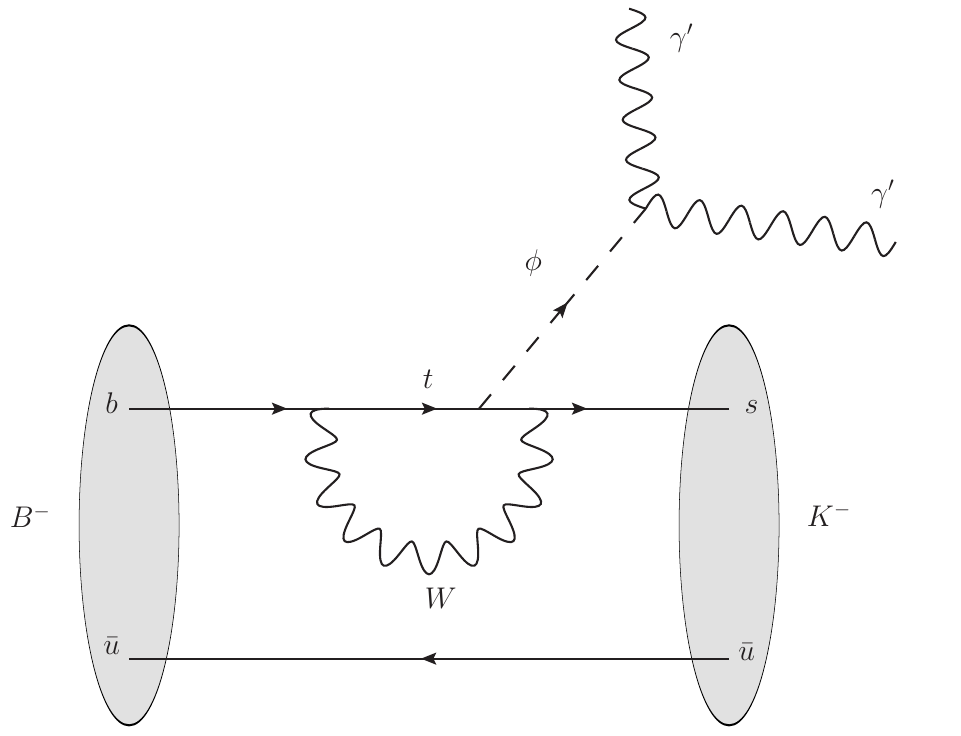}
 \includegraphics[width=0.495\textwidth]{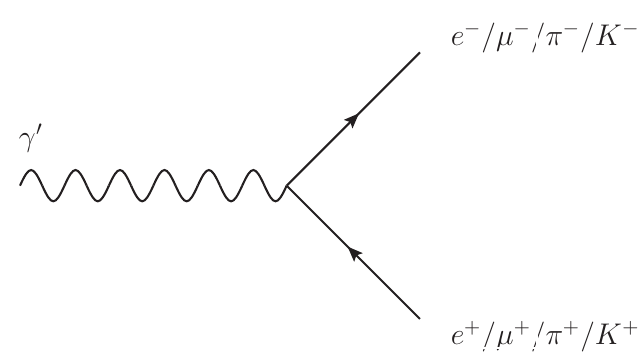}
  \caption{The Feynman diagrams for the production (left) and decay (right) of the long-lived dark photons at Belle II.
  For the production, a counterpart process for $B^+$ decays is implicitly included.
  }
  \label{fig:feynman_diagrams}
\end{figure}
We show in Fig.~\ref{fig:feynman_diagrams} the Feynman diagrams for the production and decay of the dark photons corresponding to the signature of interest.

\section{Experiment and simulation}\label{sec:experiment_simulation}

\subsection{The Belle II experiment}\label{subsec:exp}

The Belle~II experiment collects data at the SuperKEKB~\cite{Ohnishi:2013fma,Akai:2018mbz} collider at KEK in Tsukuba, Japan.
SuperKEKB collides electron and positron beams at 7 GeV and 4 GeV, respectively.
The resulting center-of-mass (CM) energy corresponds to the mas of the $\Upsilon(4S)$ resonance, which decays promptly to two $B$-mesons.
The Belle~II detector~\cite{Belle-II:2010dht,Belle-II:2018jsg} is a magnetic spectrometer of a cylindrical structure placed around the beam IP, covering over 90\% of the full $4\pi$ solid angle. 
The detector consists of several subdetectors.
Closest to the interaction point are two layers of silicon-pixel detectors, surrounded by four layers of silicon strip detectors.
These are used to track charged particles and measure decay-vertex positions with ${\cal O}(\SI{}{\micro\meter})$ precision~\cite{Belle-IISVD:2022kkx,Belle-IISVD:2022upf}.  
Outside the vertex detectors is a helium-based small-cell drift chamber, which functions as the main tracking device and measures charged-particle momenta in 1.5~T magnetic field provided by a superconducting solenoid.
Charged hadrons (e.g.~$\pi, K$, and $p$) are identified mainly by endcap and barrel Cherenkov devices located outside the drift chamber, which are based on ring imaging and photon arrival times. 
Typical efficiencies for $K$ and $\pi$ identification are about $90\%$, while the rate for a $\pi$ faking a $K$ or vice versa is at the percent level~\cite{Sandilya:2016rpm,Kaliyar:2022edg,Belle-II:2010dht}.
Outside the Cherenkov devices, covering both barrel and endcap regions, is the electromagnetic calorimeter, which consists of 8736 CsI(Tl) crystals with a depth of about 16 radiation lengths.
The calorimeter measures photon energies and provides electron identification with a typical efficiency of about 90\% and a fake rate of under $1\%$ in most kinematic range~\cite{Charged:2895,Milesi:2020esq}.
Outside the superconducting coil that encloses the calorimeter is the muon and $K_L^0$ identification system, consisting of detectors placed between the magnetic flux return iron plates.
The typical efficiency for muon identification is about 95\%, with a fake rate of a few percent~\cite{Charged:2895,Milesi:2020esq}.

\subsection{Signal-candidate reconstruction}\label{subsec:signal}

In what follows we describe the selection criteria that would likely be used in a future data analysis to suppress backgrounds and search for the signal. 
The kinematic region defined by these cuts is referred to as the signal region.
As will be described below in Sec.~\ref{subsec:bgd}, the background yield in the signal region is expected to be small.
Consequently, the search method involves estimation of the expected background yield followed by determination of the statistical consistency of the observed signal-region yield with the background estimate.
In case of consistency, one computes limits on the model parameter space.
Conversely, an observed yield significantly larger than the expected background yield implies discovery of a signal and is followed by further studies of this signal. 

Our signal process is $B^-\to K^- \phi$, with the dark scalar promptly undergoing the decay $\phi\to \gamma'\gamma'$. 
Each dark photon is reconstructed from its decay to two charged particles, which leave visible tracks in the detector and hence are denoted $t^+ t^-$.
The $t^+ t^-$ final state is primarily $e^+e^-$, $\mu^+\mu^-$, and $\pi^+\pi^-$, with a smaller $K^+ K^-$ contribution (see details in Fig.~\ref{fig:br}).
We do not consider the decay $\gamma'\to p \bar p$, which is kinematically forbidden for almost all of the relevant $m_{\gamma'}$ range given the production mechanism considered here.

We focus on the case in which the dark photon is long lived, so that its decay position is visibly displaced from its production point, yet is inside the tracking volume of Belle~II.
Thus, the two tracks from each $\gamma'\to t^+ t^-$ decay form a DV. 
Requiring the two DVs to be significantly displaced from the interaction point of the collider beams strongly suppresses background from promptly produced tracks~\cite{Lee:2018pag}. 
In our estimates (Sec.~\ref{subsec:simulation}) we take the displacement requirement to be $r_{\text{DV}}>1$~cm in the plane transverse to the collider beams, as in Ref.~\cite{BaBar:2015jvu}.
For consistency with the signal hypothesis and suppression of combinatorial, material-interaction, and $K_L^0$-decay background (see Sec.~\ref{subsec:bgd}), one will further require that the angle $\alpha$ between the $t^+t^-$ momentum measured at the DV and the vector between the interaction point and the DV be small.
For example, in Ref.~\cite{BaBar:2015jvu} the requirement used was $\alpha < 0.01$. 
The $t^+t^-$ invariant $m_{t^+t^-}$ will be required to be large enough, e.g.~$m_{t^+t^-}>20$~MeV, to suppress photon-conversion background.
In addition, one will require the difference $\Delta m$ between the invariant masses of the two dark-photon candidates to be smaller than 3 or 4 times its resolution $\sigma_{\Delta m}$.
The typical invariant-mass resolution is a few MeV for $r_{\rm DV}$ of order centimeters from the interaction point (see, e.g.~Fig.~6 of Ref~\cite{Uno:2018kbl} for the $K_S^0$ invariant-mass resolution in early Belle~II data) and degrades slowly with increasing $r_{\rm DV}$ (see Chapter 9 of Ref.~\cite{Kolanoski:2020ksk}).

Subsequently, standard selections used for $B$-meson reconstruction will be applied using the energy difference $\Delta E= E_B - \sqrt{s}/2$ and the beam-constrained mass $M_{bc}=\sqrt{s/4 - p_B^2}$.
Here, $E_B$ and $p_B$ are the measured energy and momentum, respectively, of the $B$ candidate in the CM frame of the $e^+e^-$ collision, and $s$ is the average squared CM energy of the collision. 
The value of $s$ and the boost vector from the laboratory to the CM frame are known from calibration.
Signal events are distributed as a peak around $\Delta E=0$ and $M_{bc}$ equal to the known mass of the $B^+$ meson~\cite{ParticleDataGroup:2020ssz}. 
The typical resolutions of these variables are $\sigma_{\Delta E} \approx 25$~MeV and $\sigma_{M_{bc}} \approx 2.5$~MeV for prompt decays, with slow degradation as a function of $r_{\rm DV}$. 
We expect the cuts on $\Delta E$ and $M_{bc}$ to be about 3 or 4 times their resolutions around the expected values for signal.

\subsection{Potential background sources}\label{subsec:bgd}

We consider two general types of background:
peaking background, which has a final-state signature that is similar to that of signal, and combinatorial background, which arises from random combinations of particles that meet the event-selection criteria by coincidence.

Peaking background arises from $B^+\to K^+\pi^0$, $B^+\to K^+\eta$, or $B^+\to K^+\eta'$ with the $\pi^0$, $\eta$, or $\eta'$ decaying to two photons that undergo conversion to $e^+e^-$ in detector material to form two DVs.
In the case of the $\eta$, this background can be effectively suppressed by disregarding the $\gamma'\to e^+e^-$ channel when the scalar-candidate mass $m_\phi$ is close to $ m_\eta\approx 0.5$~GeV. 
For $m_\phi\sim m_{\pi^0}\approx 0.135$~GeV, $\gamma'\to e^+e^-$ is the only kinematically allowed channel and must be used.
This inevitably leads to reduced sensitivity for $m_\phi$ within about 10~MeV of $m_{\pi^0}$.  
We note that our sensitivity estimates are given for a sample of $m_\phi$ values that are far from the masses of the $\pi^0$, $\eta$, and $\eta'$.
Later in this subsection we discuss additional measures to suppress photon-conversion background from either peaking or combinatorial background.

The decay $B^+\to K^+ K_S^0 K_S^0$ with $K_S^0 \to \pi^+\pi^-$ also constitutes peaking background for the $\gamma' \to \pi^+\pi^-$ signal mode.
This background is suppressed very effectively by rejecting events in which $m_{t^+t^-}$ is close to the mass of the $K_S^0$.
In the other signal modes, this background contribution is very strongly suppressed by particle-identification criteria and, if needed, can be further suppressed by the above $m_{t^+t^-}$ cut, taking the track masses to be that of the pion.
This approach was taken, e.g.~in Ref.~\cite{BaBar:2015jvu}.
We note also that for $m_{\gamma'}\sim m_{K_S^0}$, the dark photon decays dominantly to lepton pairs, so the impact of the $\gamma' \to \pi^+\pi^-$ channel on the sensitivity is small and is not considered in our results.

After removing the peaking and the photon-conversions backgrounds, the dominant background is combinatorial.
We estimate the abundance of this background in three steps. 
First, we consider the combinatorial background observed in BABAR and Belle analyses of related final states from prompt decays.
Second, we consider the background-reduction impact of requiring DVs for the $\gamma'$ decays.
In the third step, we discuss the impact of having two $\gamma'$ candidates in the signature.

In the first step, we consider the combinatorial background separately for leptonic and hadronic decays of the dark photons. 
For leptonic decays, one would like to use studies of $B$ decays to a kaon and four leptons. 
However, lacking published results with this final state, we instead consider BABAR~\cite{BaBar:2008jdv} and Belle~\cite{BELLE:2019xld} studies of $B^+\to K^+\ell^+\ell^-$. 
Plots of $M_{bc}$ for these studies exhibit 10-30 combinatorial-background events per ${\rm ab}^{-1}$ under the signal peak.
Relative to $B^+\to K^+\ell^+\ell^-$, our signal decay contains two additional leptons, yet softer particles overall. 
These differences, respectively, lead to a reduction and an increase in the expected background level, which we take to approximately cancel out.
For hadronic final states, we estimate a background rate of about 500 events per ${\rm ab}^{-1}$ from the BABAR~\cite{BaBar:2006qhm} study of $B^+\to K^{*0}\pi^+\pi^-$.
Since that study involved an 80-MeV-wide cut on the invariant mass of the $K^{*0}\to K^-\pi^+$ candidate, this estimate should be multiplied by roughly $\frac14 M_B/ 80~\rm MeV\approx 17$, where $\frac14 M_B$ is a rough estimate for the average invariant mass of two light particles in a 4-body decay. 
Thus, the resulting background level is about 8500 events per ${\rm ab}^{-1}$.
For events in which one $\gamma'$ decays leptonically and the other decays hadronically, one can expect the background to be the geometric average of the fully leptonic and fully hadronic final states, i.e.~around 400 events per ${\rm ab}^{-1}$.

In the second step, we note that the background level is greatly reduced by the requirement that the 4 displaced tracks originate from two DVs.
For general discussions and examples of this background-suppression effect, see e.g.~Refs.~\cite{Lee:2018pag,BaBar:2015jvu,Belle:2013ytx,Dib:2019tuj,Dey:2020juy}. 
Background sources that give rise to a DV includes mostly true DVs from particle decays, with additional contributions from particle-material interactions and accidental spatial crossing of charged-particle tracks.
We discuss these background sources in more detail in what follows.

True DVs are created in large numbers from the decays $K_S^0\to \pi^+\pi^-$ and $\Lambda\to p\pi^-$.
Such background is effectively rejected with $m_{t^+t^-}$ cuts, as discussed above. 
A smaller source of true-DV background is the decays $K_L^0\to \pi^+\pi^-\pi^0$ and the $\mathcal{O}(\%)$ of $K_L^0\to \pi^\pm \ell^\mp \nu$ decays that survive the particle-identification requirements. 
Given the long lifetime of the $K_L$, $c\tau_{K_L^0}\approx 15$~m, and its typical boost factor $\gamma_{K_L^0}\beta_{K_L^0}\sim 1$, only a few percent of $K_L^0$ mesons decay in the detector's tracking volume. 
Being three-body, these decays do not peak in the $m_{t^+t^-}$ mass, so they cannot be rejected by cutting on this variable.
However, for the same reason, they are effectively suppressed by the $\alpha$ requirement (see Sec.~\ref{subsec:signal}).

DVs from particle-material interactions involve mainly photon conversions in the $\gamma'\to e^+e^-$ channel and hadronic interactions that mostly produce pions and eject protons or nuclear fragments.
Accurate estimation of the contribution of material-interaction background to the final analysis requires full detector simulation with an event-sample size similar to that of the experimental sample, which would be beyond the scope of the current study.
Therefore, we take aggressive measures to suppress the photon-conversion backgrounds, and briefly discuss the potential application of such methods to hadronic-interaction background as well.
We note that following full-simulation study as part of the eventual experimental search, these requirements will be better tuned to the actual needs of the analysis.

Material-interaction background can be suppressed by vetoing DVs that are inside or near dense detector material layers.
Mapping the material in sufficient detail is a technical challenge, which may be avoided altogether by requiring DVs to be inside the gaseous volume of the drift chamber. 
This approach was taken, e.g.~in Ref.~\cite{Dib:2019tuj}.
At Belle~II, this corresponds to requiring the radial position of each DV to satisfy $r_{\rm DV}>16.8$~cm.
In our study, we apply this requirement only in the $e^+e^-$ channel for $m_{e^+e^-}<100$~MeV, to suppress photon-conversion background.
This requirement can be applied also for larger masses and other final states if this is determined needed by detailed detector simulation.
It is important to note that the requirement $r_{\rm DV}>16.8$~cm leads to reduced sensitivity mostly at larger values of $\epsilon$, which are probed with other methods, particularly prompt dark-photon decays. 
Material interactions occur also in the detector gas, but at a rate reduced by a factor of $\mathcal{O}(10^2)$ per DV.
Nonetheless, to aggressively suppress photon-conversion background, we apply the cut $m_{e^+e^-}>20$~MeV in the $\gamma'\to e^+e^-$ channel.
Minimal requirements on $m_{t^+t^-}$ can also be considered for other channels following full detector simulation.

Displaced-vertex background may also arise from accidental spatial crossings of tracks. 
Since the majority of tracks originate from close to the collider interaction point, this background is suppressed by requiring that the tracks forming the displaced vertex be inconsistent with originating from near the IP. 
Furthermore, for DVs that are outside the innermost detector layer, it is required that the tracks should not have detector hits at radii smaller than that of the DV.

In the third background-assessment step, we note that while the probability for occurrence of a single displaced vertex in background events is small, the probability for two such vertices is much smaller still.
Additional background suppression arises from requiring the two $\gamma'$ candidates to have consistent invariant masses (see e.g.~Ref.~\cite{BaBar:2012bkw}). 
Furthermore, the presence of two distinct vertices in the signal decay provides additional handles on background suppression if needed. 
For example, to further suppress photon-conversion background, one can allow only one of the two dark photons to decay via the di-electron channel. 
A similar criterion can be applied in the case of di-pion vertices to further suppress background from $K_S^0\to \pi^+\pi^-$ with a badly mis-measured invariant mass and from $K_L^0\to\pi^+\pi^-\pi^0$.
Similarly, if the background for two hadronic DVs is determined to be too high in the final experimental analysis, these states can be discarded, requiring that at least one DV be leptonic.
In our study we do not take such measures.

Starting from the initial background estimation of the first step and applying the background-suppression methods of the second and third steps, we conclude that the level of background can be reduced to the sub-event level without a large loss of signal efficiency, even with the full dataset of Belle~II.

The above discusison is our a-priori estimation of the background. 
In the future data analysis, the expected number of background events will be more robustly estimated using a data-driven method. 
Generally, this involves counting the observed event yields in control regions designed to have negligible signal efficiency while containing many more background events than in the signal region.
For example, requiring $m_{t^+t^-}$ to be around the $K_S^0$ mass or below about 20~MeV enhances the $K_S^0$ and photon-conversion background, respectively. 
A control region defined by, e.g.~$10 \sigma_{\Delta m} < \Delta m < 20 \sigma_{\Delta m}$ can be used to enhance background from material-interaction, $K_L^0$, and random-combination DVs. 
Another control region, defined by $10 \sigma_{M_{bc}} < M_{bc} < 20 \sigma_{M_{bc}}$, can be used to study all sources of background.
From the observed event yields in the control regions one can estimate the background yields in the signal region using simulation.
The procedure can be validated by using validation regions, defined similarly to the control regions but with different numerical values of the cuts, e.g.~$5 \sigma_{M_{bc}} < M_{bc} < 10 \sigma_{M_{bc}}$.

\subsection{Simulation procedure}\label{subsec:simulation}

In order to perform numerical simulation of the signal process described in Sec.~\ref{subsec:signal}, we employ the Monte-Carlo (MC) event generator MadGraph5aMC@NLO~\cite{Alwall:2011uj,Alwall:2014hca} with the UFO model file \texttt{HAHM}\footnote{The model file is available for download at \url{https://github.com/davidrcurtin/HAHM}.}~\cite{Curtin:2013fra,Curtin:2014cca}.
Since the model entails only flavor-diagonal interactions for the dark scalar $\phi$, we introduce an effective vertex associated with the $b-s-\phi$ interaction and subsequently modify the UFO model file with FeynRules~\cite{Alloul:2013bka}.

At the operation level of the event generation we generate the process $e^+e^- \to b\bar{b}$.
The electron and positron beams have energies of 7 and 4~GeV, respectively, corresponding to a CM energy of $\sqrt{s}=10.58$~GeV. 
The bottom quark ($b$) then undergoes the decay $b \to s \phi$.
The $\phi$ decays into a pair of dark photons.
No parton-level cuts are applied.
Our simulation is operated at the quark level, while the physical process is $e^+e^- \to B^+B^-$, with $B^+ \to K^+ \phi$.
Naively, this should lead to the simulation of wrong angular distributions.
However, we set the $b$- and $s$-quarks masses to those of the $B^+$- and $K^+$ mesons, respectively, so that the $b$ quarks have very little velocity in the CM frame.
This results in the $s$ quarks and $\phi$ bosons being uniformly distributed in $\cos\theta_p^*$ (where $\theta_p^*$ is the polar angle with respect to the beams in the CM frame), as is the case in the physical process.

We perform parameter scans of the model in $m_\phi$, $m_{\gamma'}$, and $\epsilon$. 
We choose five representative values of $m_\phi$ ranging from 0.1 GeV to 4.0 GeV.
For each value of $m_\phi$, we simulate samples with different values of $m_{\gamma'}$ from 0.02 GeV to $m_\phi/2$.
The MadGraph5 simulation outputs LHE files~\cite{Alwall:2006yp} containing the signal-event information.
We apply the Python-based tool Pylhe~\cite{pylhe} to read in these files and then perform further analysis and computation.

For each simulated sample, we use the kinematics of the simulated events to calculate the expected number of observed signal events at the Belle~II experiment for different values of $\epsilon$:
\begin{equation}\label{eq:Prob}
\begin{aligned}
     N_S(m_\phi, m_{\gamma'}, \epsilon) & = 2 \times N_{e^+e^- \to B^+ B^-} \times \mathcal{B}(B^+ \to K^+\phi) \times \mathcal{B}(\phi \to \gamma'\gamma')  \times \varepsilon^{\text{trk}} \\
    & \times   \sum_{t_{i,j}=e, \mu, \pi, K}~ \varepsilon_{ij}^{\text{PID}}\cdot  
 \mathcal{B}(\gamma' \to t_i^+t_i^-) \times \mathcal{B}(\gamma' \to t_j^+t_j^-) ,
\end{aligned}
\end{equation}
where $N_{e^+e^- \to B^+ B^-}=2.75\times 10^{10}$ is the predicted number of $B^+B^-$ events at Belle~II with an integrated luminosity of 50 ab$^{-1}$; 
$\mathcal{B}$ indicates a branching fraction\footnote{Recall that we take $\mathcal{B}(\phi \to \gamma' \gamma') \simeq 1$; see Sec.~\ref{sec:model}}; 
$\varepsilon^{\text{trk}}$ is the tracking efficiency, defined as the average probability of detecting both dark-photon decays in the event;
and $\varepsilon_{ij}^{\text{PID}}$ is the particle identification efficiency, defined as the probability to identifying the displaced tracks as electrons, muons, pions, or kaons according to the final state given by the indices $i,j$.

The tracking-related signal efficiency is calculated as:
\begin{equation}
    \varepsilon^{\text{trk}}=\frac{1}{N_{\text{sim}}}\sum_{k=1}^{N_{\text{sim}}}P^{\gamma '_{k1}}\,P^{\gamma '_{k2}},
\end{equation}
where $N_{\text{sim}}$ is the total number of the simulated events and $P^{\gamma '{_{k1}}}$ and $P^{\gamma '_{k2}}$ represent, respectively, the probabilities of the first and the second dark photons in the $k^{\text{th}}$ simulation event to be detected.
For event $k$, the probability for detection of dark photon $n$ is calculated as 
\begin{equation}
P^{\gamma'_{kn}} = \frac{1}{R}\int_0^{80~\text{cm}} L(r) A(z) e^{-\frac{r}{R}} \, dr,  
\end{equation}
where
$R=(p_T^{\gamma'_{kn}}/m_{\gamma'}) c\tau_{\gamma'}$ is the average transverse flight distance of the dark photon $n$ before its decay, with $p_T^{\gamma'_{kn}}$ being its simulated transverse momentum, and the lifetime $\tau_{\gamma'}$ being determined from $m_{\gamma'}$ and $\epsilon$ (see Sec.~\ref{sec:model});
and $z= r \cot\theta_p$ is the longitudinal coordinate for the decay position of the dark photon that corresponds to the radial coordinate $r$, with $\theta_p$ being the polar angle of the dark photon in the laboratory frame.
Furthermore, the function
\begin{equation}
    A(z)= \left\{ \begin{matrix}
        1 & , & -40 < z < 120~\text{cm} \\
        0 & , & \text{otherwise}
    \end{matrix}
    \right\}
\end{equation}
represents the longitudinal extent $-40 < z < 120~\text{cm}$ of the fiducial volume of the tracking volume, within which DVs can be detected.
Lastly, the function
\begin{equation}
    L(r)= \left\{ \begin{matrix}
        1 - \frac{r}{80~\text{cm}} & , & 1 < r < 80~\text{cm} \\
        0 & , & \text{otherwise}
    \end{matrix}
    \right\}
\end{equation}
corresponds to the radial extent $1 < r < 80~\text{cm}$ of the fiducial volume, our cut $r_{\text{DV}}>1$~cm\footnote{For $m_{\gamma'}< 0.1$~GeV and $\gamma'$ decays into $e^+ e^-$, the radius cut is $16.8 < r_{\text{DV}} < 80~\text{cm}$, see Sec.~\ref{subsec:exp}.}, and a linear drop in the tracking efficiency with radius.
This approximate parameterization of the fiducial volume and tracking efficiency follows Refs.~\cite{Dib:2019tuj, Dey:2020juy, Cheung:2021mol,Bertholet:2021hjl}.

\begin{figure}[t]
	\centering
	\includegraphics[width=1\textwidth]{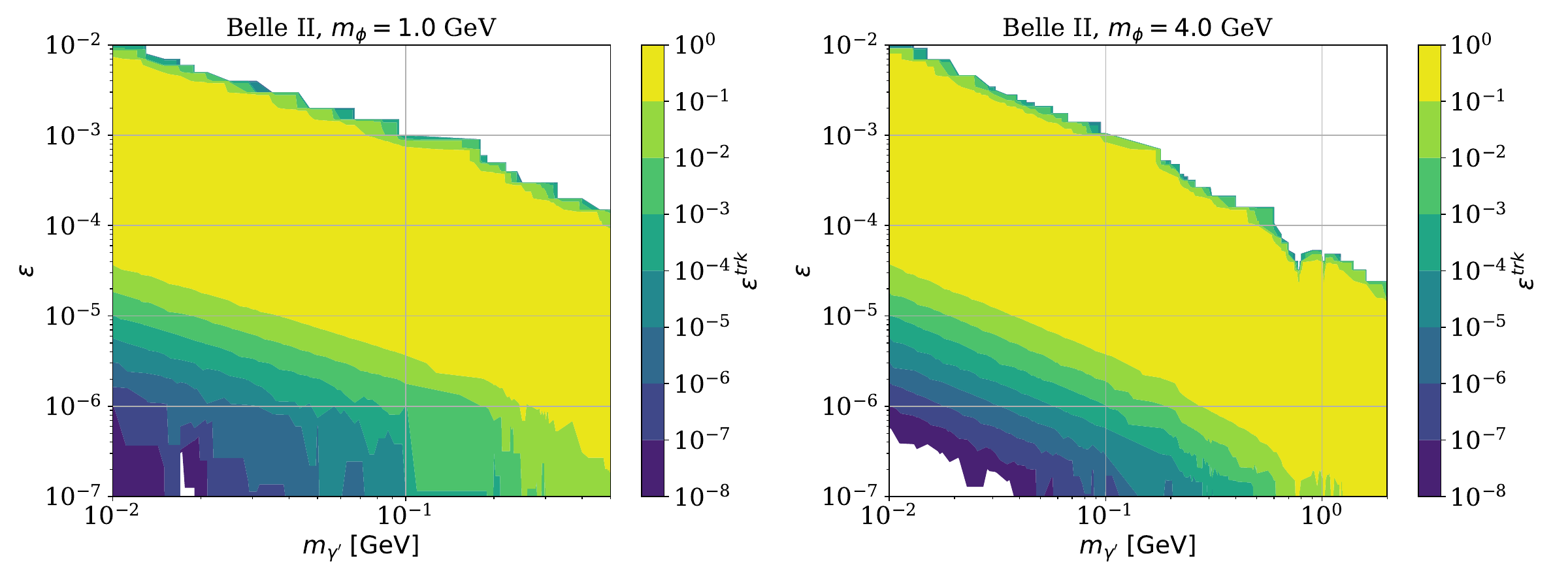}
	\includegraphics[width=0.5\textwidth]{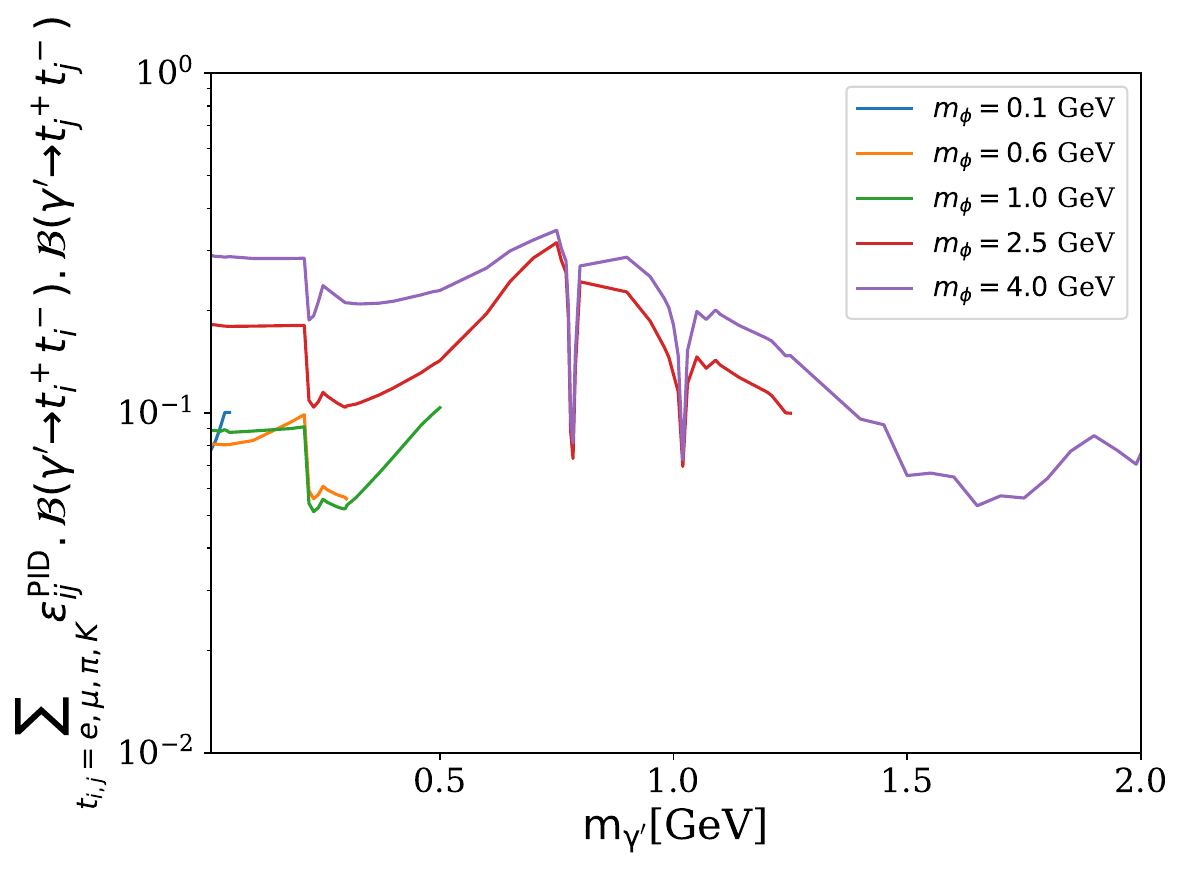}
	\caption{\textit{Upper panels}: Density plot of $\epsilon^{\text{trk}}$ shown in the plane $\epsilon$ vs.~$m_{\gamma'}$, for $m_\phi=1.0$ GeV (\textit{left}) and $m_\phi=4.0$ GeV (\textit{right}).
 The white-space parts are where $\epsilon^{\text{trk}}$ is so small that it is considered as zero by the machine.
 \textit{Lower panel}: the sum of $\varepsilon_{ij}^{\text{PID}}\cdot  \mathcal{B}(\gamma' \to t_i^+t_i^-) \cdot \mathcal{B}(\gamma' \to t_j^+t_j^-)$ over all the dark-photon final-state combinations, as functions of $m_{\gamma'}$, for dark-Higgs masses of 0.1, 0.6, 1.0, 2.5, and 4.0 GeV.
 }
 \label{fig:epsilonTRK_epsilonPID_BR_BR}
\end{figure}

We present in the upper panels of Fig.~\ref{fig:epsilonTRK_epsilonPID_BR_BR} density maps of $\epsilon^{\text{trk}}$ in terms of the kinetic mixing parameter $\epsilon$ and the dark-photon mass $m_{\gamma'}$, for two benchmark dark-Higgs masses $m_{\phi}=1.0$ and 4.0 GeV.
One observes that in large parts of the parameter space, $\epsilon^{\text{trk}}$ is of order 10\%.
We note that the blank space in these plots is where the computed value of $\epsilon^{\text{trk}}$ is below the machine precision and thus considered as zero.

The particle-identification efficiency $\varepsilon_{ij}^{\text{PID}}$ is calculated with a separate simulation.
We use the EVTGEN~\cite{evtgen} event generator to produce the signal decays, employing the following models within EVTGEN.
The decay $B^{+}\rightarrow K^{+} \phi$ is produced with the PHSP phase-space model.
The decay $\phi \rightarrow \gamma' \gamma'$ is generated with the SVV\_HELAMP model, with either the longitudinal helicity amplitude $H_0$ being non-zero or the two transverse amplitudes $H_\pm$ being non-zero and equal with a 0 relative phase.
The decays of the dark photon to two leptons or two hadrons are produced with the VLL and VSS models, respectively.
For each set of $m_\phi$ and $m_{\gamma'}$ values we produce a sample of $10^5$ events.
We determine the particle-identification efficiency for each charged particle in each event based on Figs.~25, 23, and~28 of Ref.~\cite{Belle2physbook}. 
For kaons and pions, we take the efficiency to be 90\% if the particle is within the angular acceptance range drift-chamber, $17^{\circ}<\theta_p<150^{\circ}$. 
For leptons, we simplify the $\theta_p$- and momentum-dependence of the efficiency extracted from Ref.~\cite{Belle2physbook} and report the result in  Tables~\ref{table:eff_eid} and~\ref{table:eff_muid} in Appendix~\ref{app:efficiencies_pid}.
The event-level efficiency is the product of the efficiencies for the five tracks.
For each event sample, the total particle-identification efficiency $\varepsilon_{ij}^{\text{PID}}$ is the average event-level efficiency of the sample.

This procedure does not account for impact of the dark-photon decay position on the particle-identification efficiency, and is hence imprecise. 
This simplification is necessary within the scope of this work, since particle-identification efficiencies for displaced particles are not publicly available at this time. 
Since the dedicated particle-identification detectors (the Cherenkov devices, the calorimeter, and the muon system) are all outside the drift chamber, one expects our procedure to somewhat underestimate $\varepsilon_{ij}^{\text{PID}}$.
This is because a charged particles produced at a DV is closer to the particle-identification detectors it is flying toward, and hence has a larger probability of hitting it.

In Figs.~\ref{fig:eff_figs_Kpluseeee}-\ref{fig:eff_figs_KplusKKKK} of Appendix~\ref{app:efficiencies_pid} we show the value of $\varepsilon_{ij}^{\text{PID}}$ for each final state and for different values of $m_{\phi}$ and $m_{\gamma'}$.

Finally, in the lower panel of Fig.~\ref{fig:epsilonTRK_epsilonPID_BR_BR}, we present the sum $\displaystyle \sum_{t_{i,j}=e, \mu, \pi, K}~ \varepsilon_{ij}^{\text{PID}}\cdot  \mathcal{B}(\gamma' \to t_i^+t_i^-) \cdot \mathcal{B}(\gamma' \to t_j^+t_j^-)$ (see Eq.~\eqref{eq:Prob}) as functions of the dark-photon mass $m_{\gamma'}$, for five representative dark-Higgs masses $m_\phi=0.1, 0.6, 1.0, 2.5,$ and 4.0 GeV.
Comparing this plot with Fig.~\ref{fig:br}, we conclude that the dominant factors in $\displaystyle \sum_{t_{i,j}=e, \mu, \pi, K}~ \varepsilon_{ij}^{\text{PID}}\cdot  \mathcal{B}(\gamma' \to t_i^+t_i^-) \cdot \mathcal{B}(\gamma' \to t_j^+t_j^-)$ are the branching ratios.
Further, we note that the shapes of the curves shown in the lower plot of Fig.~\ref{fig:epsilonTRK_epsilonPID_BR_BR} will also help explain certain features of the sensitivity plots we will present in the next section.

\section{Numerical results}\label{sec:results}

We proceed to present numerical results in terms of the Belle~II sensitivity for the signal.

Since the number of background events is expected to be smaller than 1 (see discussion in Sec.~\ref{subsec:bgd}), we take the edge of the parameter-space region that is excluded at the 95\% confidence level to be that for which observation of 3 signal events is expected based on Eq.~(\ref{eq:Prob}).
While we have calculated the particle-identification efficiency for both fully longitudinal and fully transverse polarizations, only the longitudinal-polarization case is used. 
The difference between the two cases is minor, and its magnitude can be gauged from the plots in Appendix~\ref{app:efficiencies_pid}.

\begin{figure}[t]
	\centering
	\includegraphics[width=0.5\textwidth]{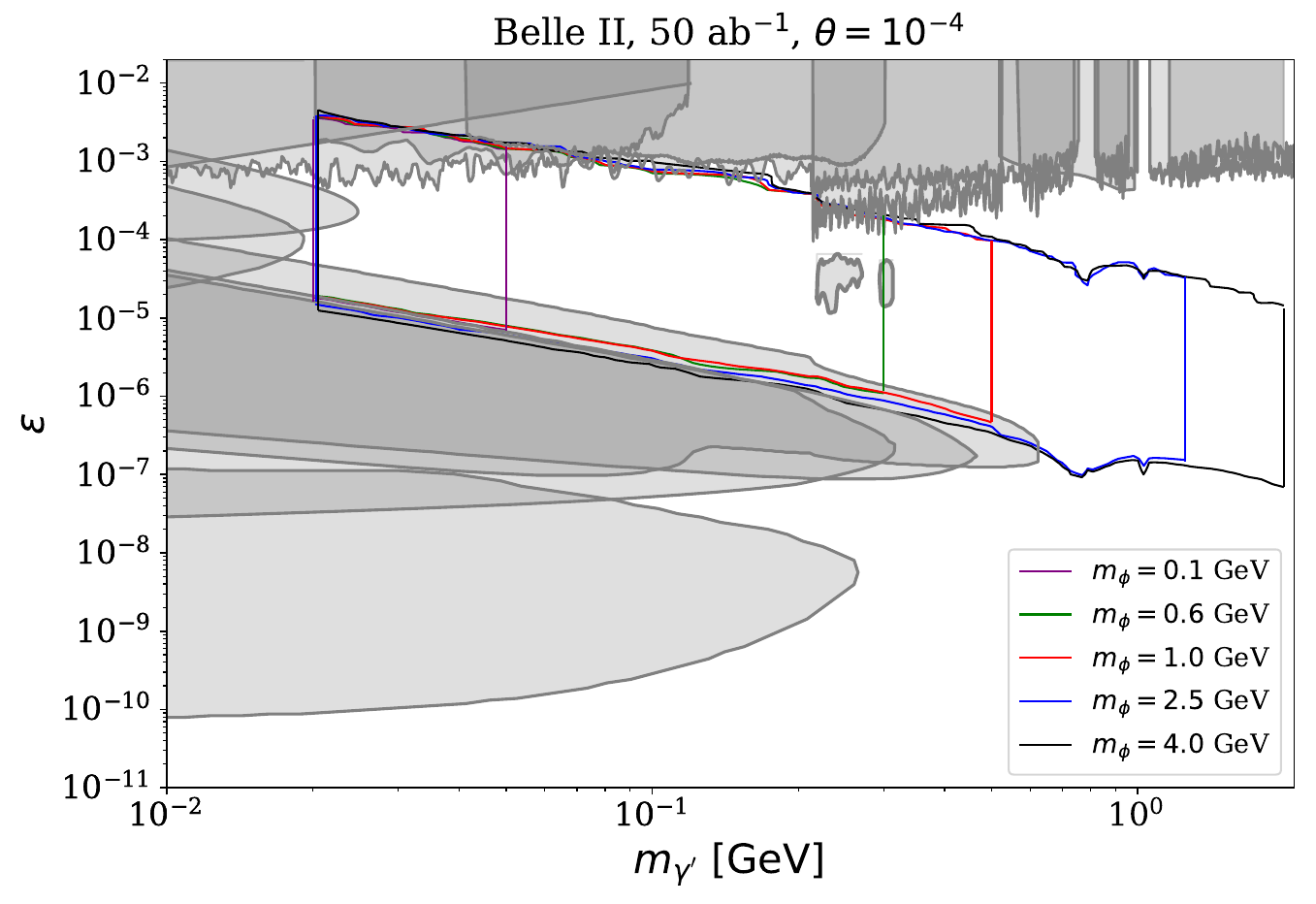}\\
	\includegraphics[width=0.4955\textwidth]{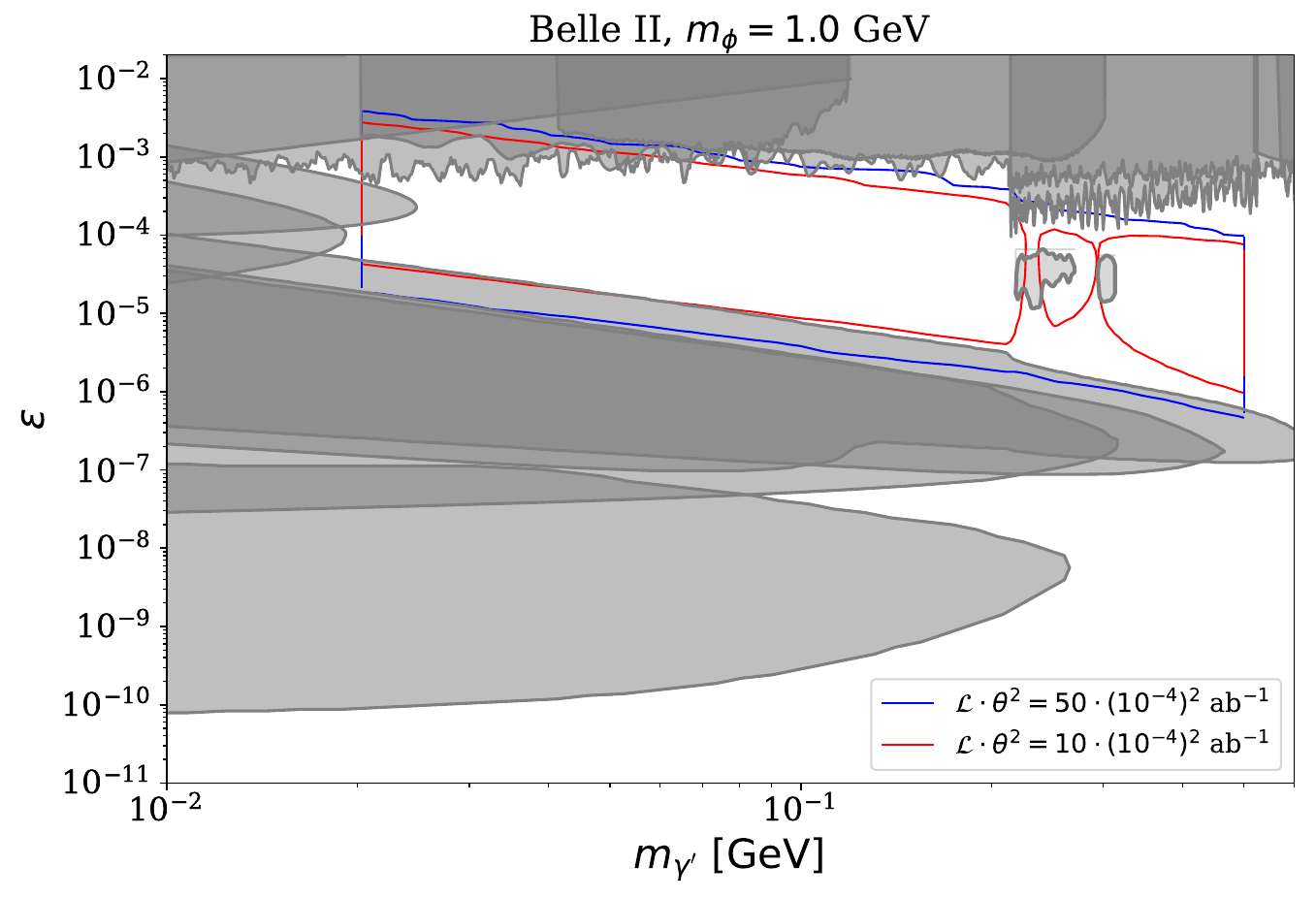}
	\includegraphics[width=0.4955\textwidth]{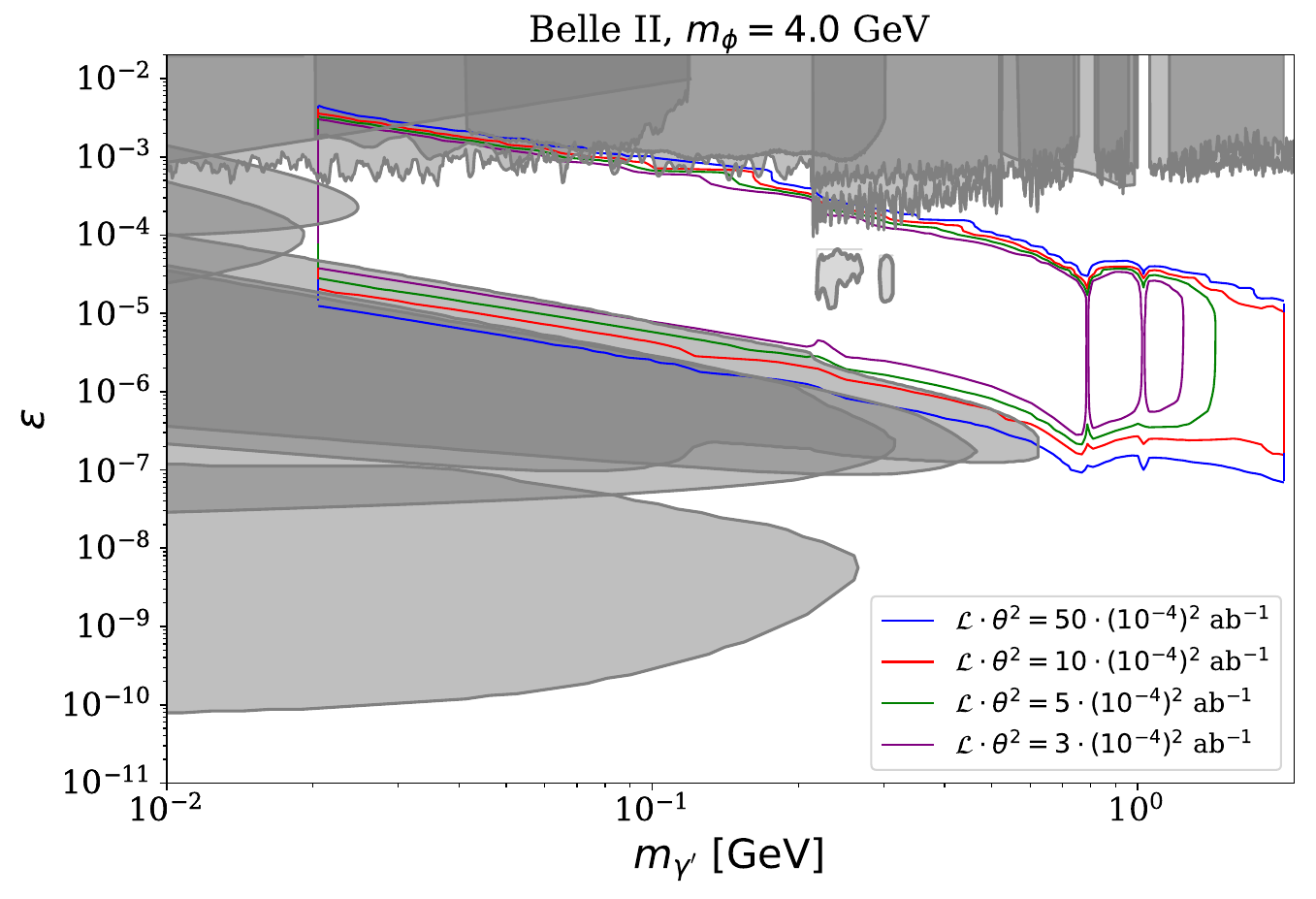}
	\caption{\textit{Upper panel:} Sensitivity results for the case of fully longitudinal polarization shown in the plane $\epsilon$ vs.~$m_{\gamma'}$, for various dark-Higgs masses with $\theta=10^{-4}$.
                \textit{Lower panels:} The same sensitivity results but for different choices of $\mathcal{L}\cdot \theta^2$, for $m_\phi=1.0$ GeV (left) and 4.0 GeV (right).
        The gray areas represent the existing limits on the massive dark photon for $m_{\gamma '} \geq \rm 10^{-2}~GeV$ from di-lepton searches at collider/fixed target experiments (A1~\cite{Merkel:2014avp}, LHCb~\cite{LHCb:2019vmc}, BaBar~\cite{BaBar:2014zli}, KLOE~\cite{KLOE-2:2011hhj,KLOE-2:2012lii,KLOE-2:2014qxg,KLOE-2:2016ydq}, and NA48/2~\cite{NA482:2015wmo}), and previous beam dump experiments: E141~\cite{Riordan:1987aw}, E137~\cite{Bjorken:1988as,Batell:2014mga,Marsicano:2018krp}, $\nu$-Cal~\cite{Blumlein:2011mv,Blumlein:2013cua}, and CHARM~\cite{Gninenko:2012eq}.
        Bounds from supernovae~\cite{Chang:2016ntp} and $(g-2)_e$~\cite{Pospelov:2008zw} are also shown together in gray.
 }
 \label{fig:sen_Long}
\end{figure}

The results are shown in Fig.~\ref{fig:sen_Long}.
In the upper panel, we overlap the sensitivity reach of Belle II with 50 ab$^{-1}$ integrated luminosity for $m_\phi=0.1, 0.6,1.0, 2.5$ and 4.0 GeV.
In this plot, the sensitivity results are presented in the ($m_{\gamma'}$, $\epsilon$) plane for the scalar mixing angle $\theta=10^{-4}$, which is allowed by the existing bounds discussed in Sec.~\ref{sec:intro}.

In the upper plot of Fig.~\ref{fig:sen_Long}, the region that can be excluded by Belle~II with $50~\text{ab}^{-1}$ is the region enclosed by the curves.
Along the top curve, the dark photon is short lived, so that not enough dark photons satisfy the minimal $r_{\text{DV}}$ cut.
Conversely, along the bottom curve the dark photon is long-lived.
We note also that the lower dark-photon mass reach is due to the cut $m_{e^+e^-}>0.02$~GeV used to suppress the photon-conversion background, and that the upper reach corresponds to the kinematic threshold $m_{\gamma'}<m_\phi/2$.

Comparing the different curves in the upper plot of Fig.~\ref{fig:sen_Long}, we observe that varying the dark-scalar mass does not have a significant impact on the sensitivity, except at the upper reach of the dark-photon mass determined by the kinematic threshold.

In the bottom plots, we consider $m_\phi=1.0$ and 4.0 GeV, respectively, showing the Belle II's sensitivity reach in the plane $\epsilon$ vs.~$m_{\gamma'}$, for different benchmarks of the combination $\mathcal{L}\cdot \theta^2$.
With $m_\phi=1.0$ GeV, only for $\mathcal{L}\cdot \theta^2=50\cdot (10^{-4})^2$ ab$^{-1}$ and $10\cdot (10^{-4})^2$ ab$^{-1}$ more than 3 signal-events are predicted in certain regions of the parameter space, while with $m_\phi=4.0$ GeV we find for all of $\mathcal{L}\cdot \theta^2=(50, 10, 5, 3) \cdot (10^{-4})^2$ ab$^{-1}$ Belle II can be sensitive to the model parameter space.
This is mainly because the dependence of Eq.~\eqref{eq:fK} on $q^2=m^2_\phi$ rendering $\Gamma(B^+\to K^+\phi)$ and hence the signal-event number grow with increasing $m_\phi$.
For values of $\mathcal{L}\cdot \theta^2$ lower than those shown, there is no sensitivity in both plots.
Naively, in Fig.~\ref{fig:sen_Long}, we expect the lower sensitivity reach in $\epsilon$ to be proportional to $(\mathcal{L}\cdot \theta^2)^{-4}$, given vanishing background.
This can be understood as follows.
Along the lower curves in Fig.~\ref{fig:sen_Long}, the dark photon is expected to be in the large-decay-length regime where, roughly speaking, its boosted decay length is much larger than the distance from its production point to the outer edges of the fiducial volume; the tracking efficiency, and hence the signal-event rate $N_S$, are proportional to $\Gamma_{\gamma'}^2$, where the power of two arises from the required observation of two dark photons in each event.
Since $\Gamma_{\gamma'} \propto \epsilon^2$ (see Eqs.~\eqref{eqn:PW_A2ll}, \eqref{eqn:PW_A2had}), we conclude that $N_S \propto\epsilon^4$.
Furthermore, $N_S\propto \mathcal{L}\cdot \theta^2$ (see Eqs.~\eqref{eqn:GammaB2Kphi}, \eqref{eq:Prob}).
As a result, decreasing $\mathcal{L}\cdot \theta^2$ by a factor of e.g.~$10^4$ leads to reduction in $N_S$ by $10^4$, which can in turn be offset by increasing $\epsilon$ by 10.
However, as we observe in e.g.~the lower left plot of Fig.~\ref{fig:sen_Long}, lowering $\mathcal{L}\cdot \theta^2$ from 50 to 10 by a factor 5, the lower sensitivity reach in $\epsilon$ is weakened by more than $5^{1/4}\sim 1.5$; this arises from the fact that along the lower curves the dark photon is not long-lived enough to be in the large-decay-length limit.

We stress that the dark-photon production rate depends on $\theta^2$ while its decay is mediated by the kinetic mixing parameter $\epsilon$.
This decoupling of the production and decay provides the advantage of expected large reach of our proposed search, compared to the minimal scenario where both the dark-photon production and decay are induced by $\epsilon$.

Fig.~\ref{fig:sen_Long} also shows as gray-shaded areas the current constraints on the dark-photon parameters, obtained from di-lepton searches conducted at colliders and fixed target experiments, including A1~\cite{Merkel:2014avp}, LHCb~\cite{LHCb:2019vmc}, BaBar~\cite{BaBar:2014zli}, KLOE~\cite{KLOE-2:2011hhj, KLOE-2:2012lii, KLOE-2:2014qxg, KLOE-2:2016ydq}, and NA48/2~\cite{NA482:2015wmo}, the beam-dump experiments E141~\cite{Riordan:1987aw}, E137~\cite{Bjorken:1988as, Batell:2014mga, Marsicano:2018krp}, $\nu$-Cal~\cite{Blumlein:2011mv, Blumlein:2013cua}, and CHARM~\cite{Gninenko:2012eq}, constraints from supernovae~\cite{Chang:2016ntp}, and the electron anomalous magnetic moment~\cite{Pospelov:2008zw}.

The combination of the existing limits and our prediction for the Belle~II sensitivity clearly demonstrates the importance of the Belle~II search proposed here.
Specifically, Fig.~\ref{fig:sen_Long} shows that the medium-$\epsilon$ regime, which is currently mostly unexcluded, falls exactly where Belle~II is the most sensitive.

Finally, we explain some features observed in Fig.~\ref{fig:sen_Long}.
For example, the lower left plot shows islands of the red curves separated at about 0.21 GeV and 0.28 GeV, which reflect the behavior of $\displaystyle \sum_{t_{i,j}=e, \mu, \pi, K}~ \varepsilon_{ij}^{\text{PID}}\cdot  \mathcal{B}(\gamma' \to t_i^+t_i^-) \cdot \mathcal{B}(\gamma' \to t_j^+t_j^-)$ as shown in Fig.~\ref{fig:epsilonTRK_epsilonPID_BR_BR}, corresponding the dimuon and di-pion thresholds, respectively.
Similarly, in the lower right panel the purple curves present also islands separated at around 0.8 and 1.0 GeV, which are due to not only the behavior of the curves displayed in Fig.~\ref{fig:epsilonTRK_epsilonPID_BR_BR} but also the sudden sharp increase of the dark-photon total decay width as plotted in the right panel of Fig.~\ref{fig:br} arising from the $\rho, \omega$, and $\phi$ resonances.
We also comment that the zigzag in the upper curves in each plot is due to insufficient statistics for the prompt regime, where only a small proportion of the generated events, those with largely boosted dark photons, contribute significantly to the computation of $N_S$.

\section{Conclusions}\label{sec:Conclusions}

In this paper we propose a displaced-vertex-based search for long-lived dark photons at the ongoing experiment Belle II, in the theoretical framework of a hidden sector with a dark scalar.
At Belle~II, $B^\pm$ mesons are pair-produced and can decay to a charged kaon $K^\pm$ and a light dark scalar $\phi$.
We consider the case of $\phi$ decaying exclusively to a pair of dark photons.
Via kinetic mixing, the dark photons subsequently decay leptonically or hadronically.
We restrict the study to the experimentally favorable final states $e^+e^-$, $\mu^+\mu^-$, $\pi^+\pi^-$, and $K^+K^-$.
We further require that both dark photons decay inside the Belle~II detector's tracking volume.

We elaborate on potential background sources and argue for their insignificance.
We perform Monte-Carlo simulations with MadGraph5 and compute the expected number of observed signal events for different values of the kinetic-mixing coefficient $\epsilon$, the dark-photon mass $m_{\gamma'}$, the  dark-scalar mass $m_\phi$, and a currently allowed value $\theta=10^{-4}$ for the mixing angle between the dark scalar and the Standard-Model Higgs.
In this simulation, we implement the displaced-tracking efficiency as a linear function of the transverse distance of the dark-photon decay position from the interaction point.
Furthermore, using the EVTGEN event generator and published information, we incorporate the particle-identification efficiency and its dependence on the final-state particles and their kinematics.

We report the sensitivity reach of our proposed search in terms of the region in $\epsilon$ vs.~$m_{\gamma'}$ that Belle~II can exclude at 95\% confidence level with an integrated luminosity of $\mathcal{L}=50~\text{ab}^{-1}$. 
Given the lack of background, this region is taken to be that for which at least three signal events would be observed. 
These bounds are calculated for five benchmark values of the $m_\phi$. 
We note that reduced sensitivity is expected for $m_\phi$ within about 10~MeV of the $\pi^0$ mass of $135$~MeV, due to background from $B^+\to K^+\pi^0$, $\pi^0\to\gamma\gamma$, with the photons undergoing conversion to $e^+e^-$ in detector material.
Further, additional sensitivity plots are shown for various values of $\mathcal{L}\cdot \theta^2$, for $m_\phi=1.0$ and 4.0 GeV.
Our results show that the search we propose uniquely probes a large, unexcluded region.

\section*{Acknowledgment}
A.~S.~is supported by grants from Israel Science Foundation, United States-Israel Binational Science Fund, Tel Aviv University, and EU Horizon 2020.  
Yk.~K.~and Yj.~K.~are supported by the National Research Foundation of Korea grant NRF-2022R1A2C1003993.
A.~S.~and Yj.~K.~acknowledge support from Tel Aviv-Yonsei Exchange Program.
K.C.~and C.J.O.~are supported by MoST under Grant no.~110-2112-M-007-017-MY3.

\appendix
\section{Detector efficiencies for different final-state particles}\label{app:efficiencies_pid}

\begin{figure}[htbp]
    \centering
    \includegraphics[width=0.45\linewidth]{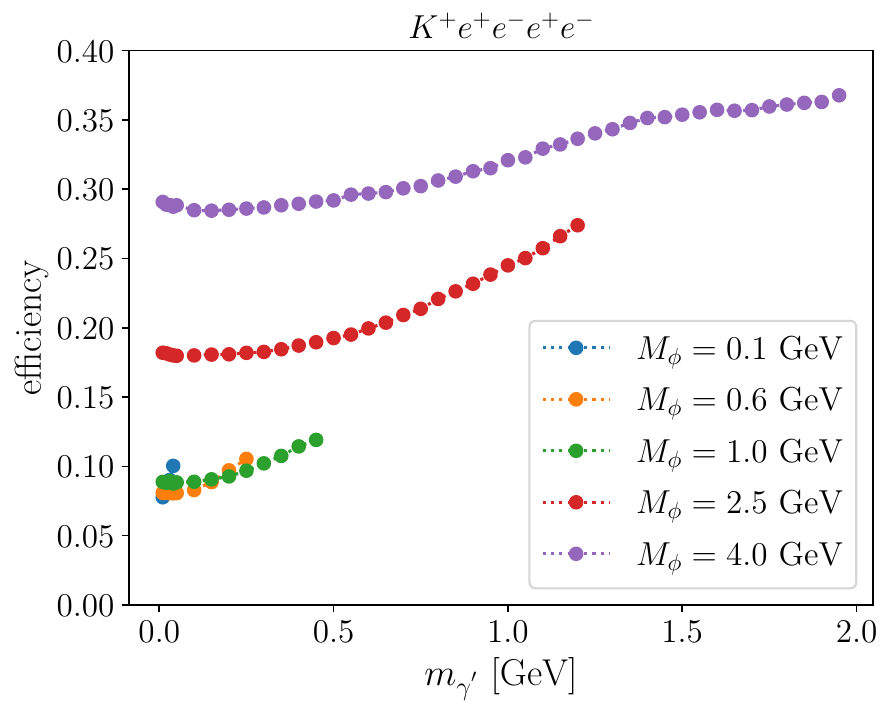}
    \includegraphics[width=0.45\linewidth]{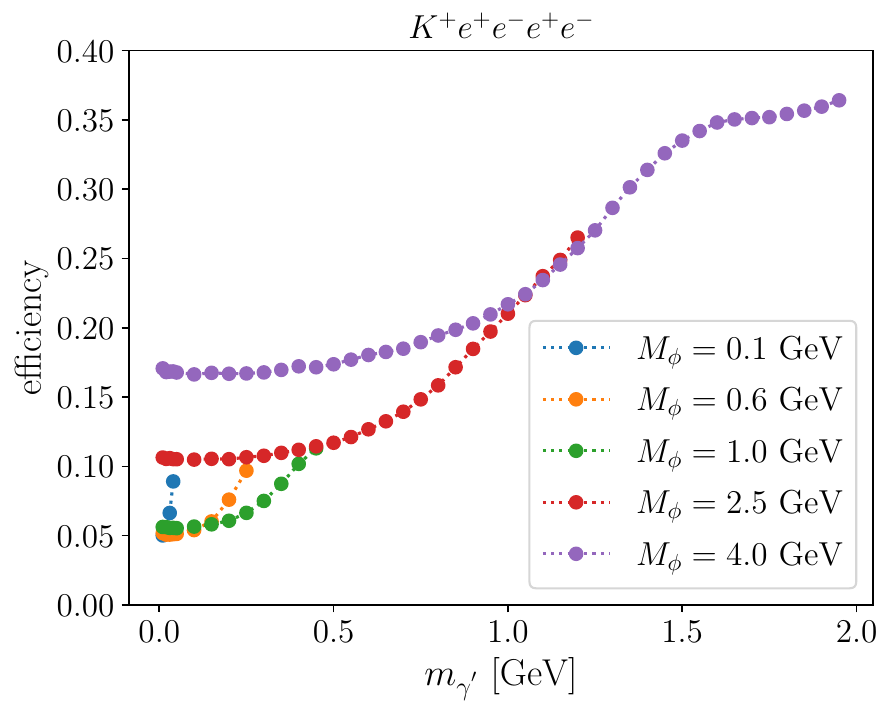}
    \caption{Estimated particle-identification efficiency of the $K^{+}e^{+}e^{-}e^{+}e^{-}$ final state with respect to the dark scalar mass and the dark-photon mass.
    The left figure shows results from fully longitudinal amplitude events and the right figure are for fully transverse amplitude events.}
    \label{fig:eff_figs_Kpluseeee}
\end{figure}

\begin{figure}[htbp]
    \centering
    \includegraphics[width=0.45\linewidth]{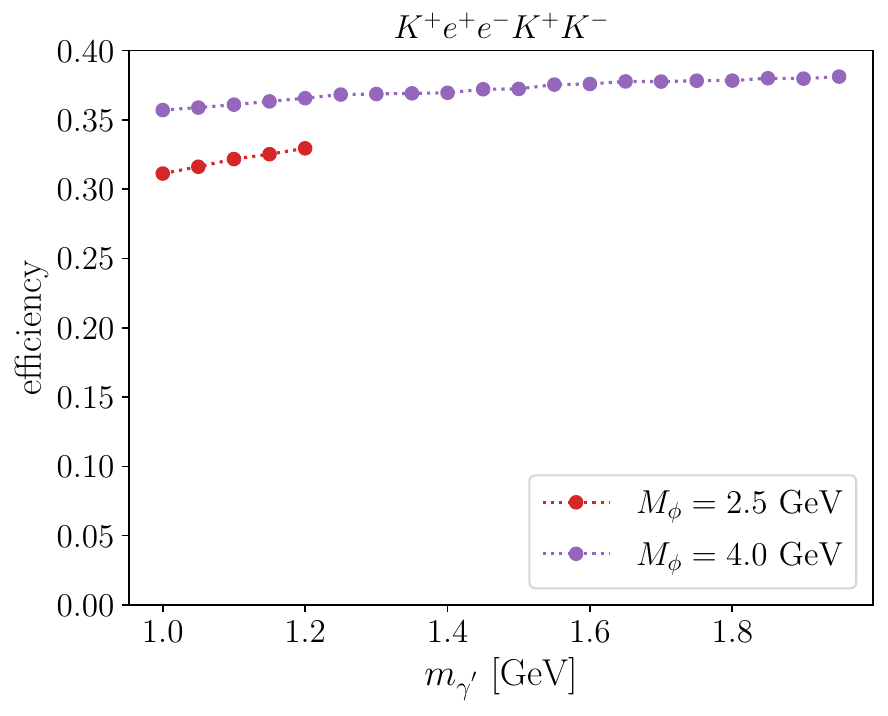}
    \includegraphics[width=0.45\linewidth]{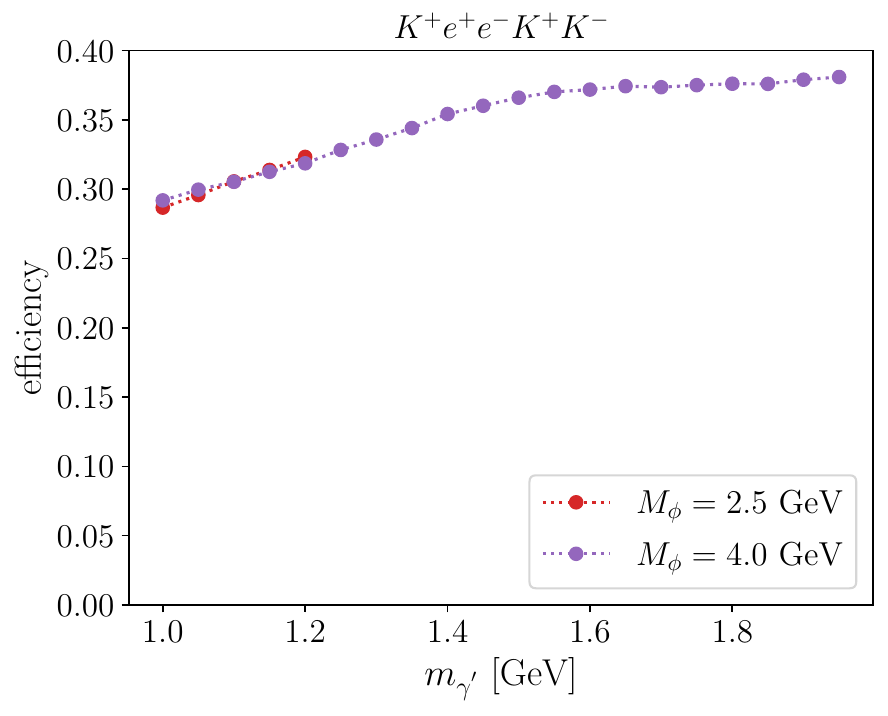}
    \caption{The same format as in  Fig.~\ref{fig:eff_figs_Kpluseeee} but for the $K^{+}e^{+}e^{-}K^{+}K^{-}$ final state.}
    \label{fig:eff_figs_KpluseeKK}
\end{figure}

\begin{figure}[htbp]
    \centering
    \includegraphics[width=0.45\linewidth]{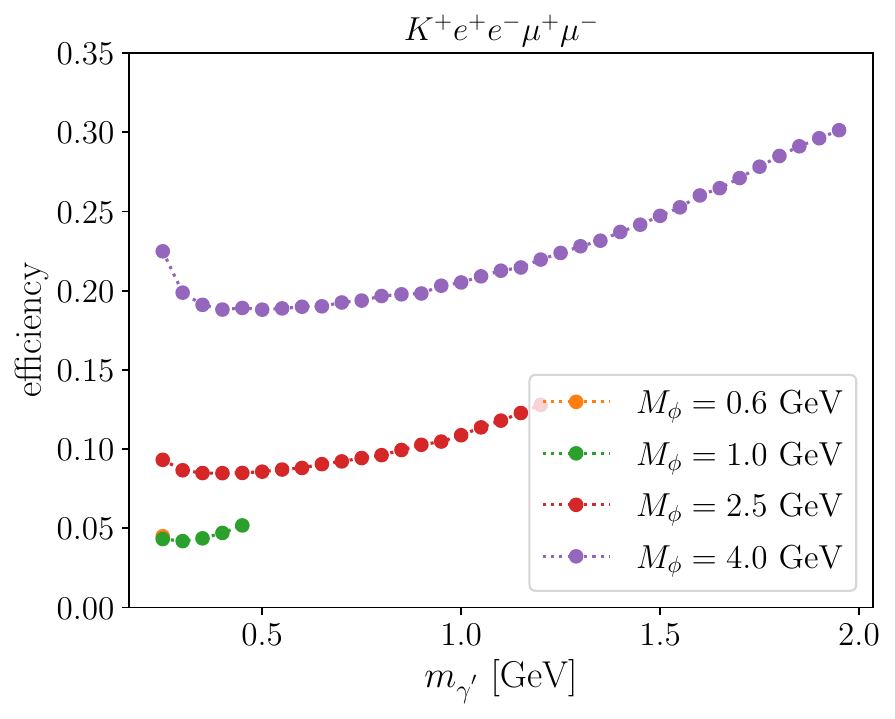}
    \includegraphics[width=0.45\linewidth]{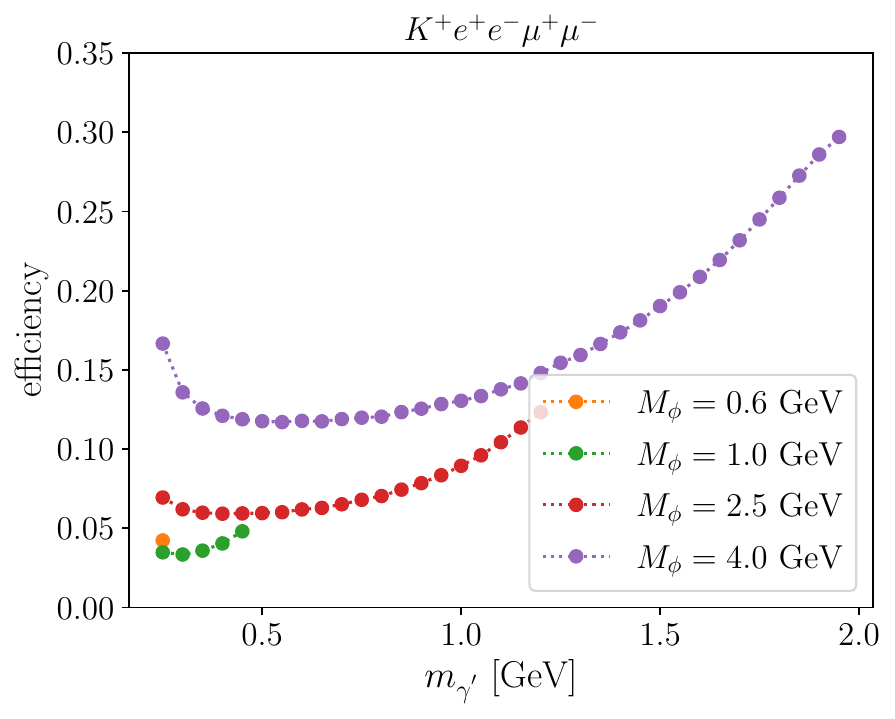}
    \caption{The same format as in  Fig.~\ref{fig:eff_figs_Kpluseeee} but for the $K^{+}e^{+}e^{-}\mu^{+}\mu^{-}$ final state.
    }
    \label{fig:eff_figs_Kpluseemm}
\end{figure}

\begin{figure}[htbp]
    \centering
    \includegraphics[width=0.45\linewidth]{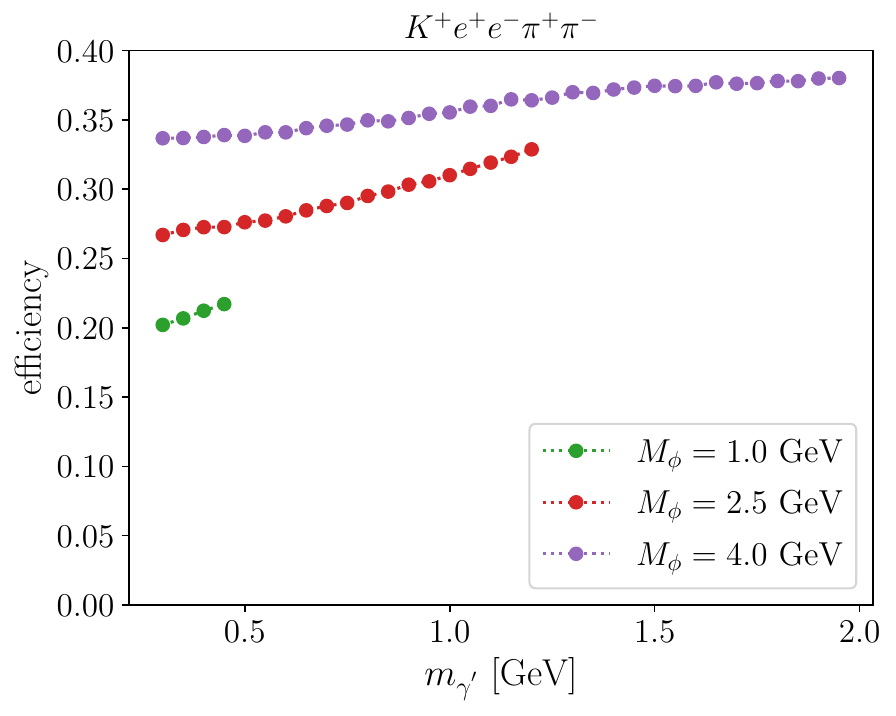}
    \includegraphics[width=0.45\linewidth]{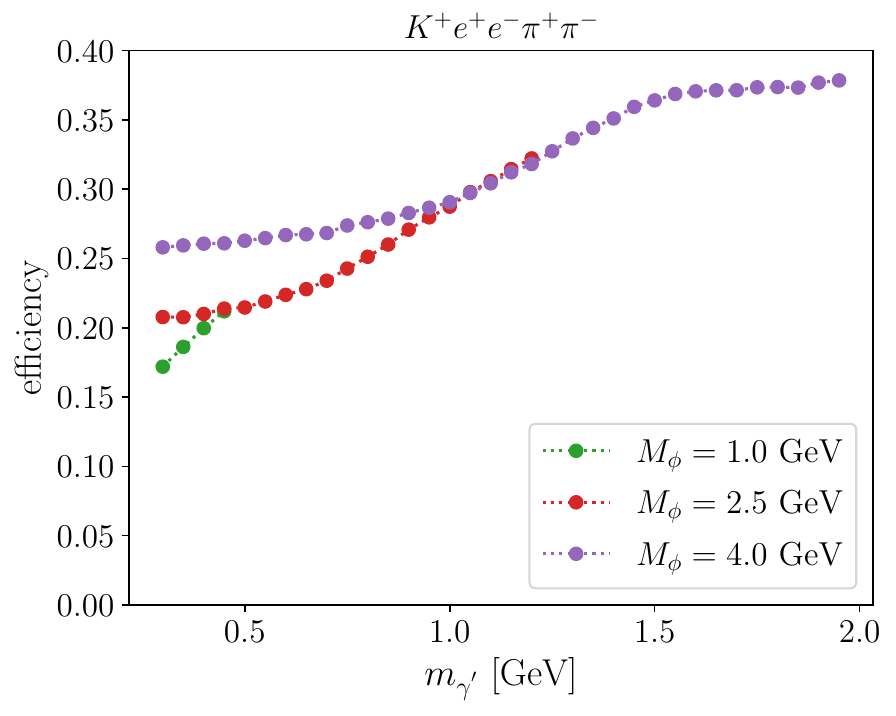}
    \caption{The same format as in  Fig.~\ref{fig:eff_figs_Kpluseeee} but for the $K^{+}e^{+}e^{-}\pi^{+}\pi^{-}$ final state.
    }
    \label{fig:eff_figs_Kpluseepp}
\end{figure}

\begin{figure}[htbp]
    \centering
    \includegraphics[width=0.45\linewidth]{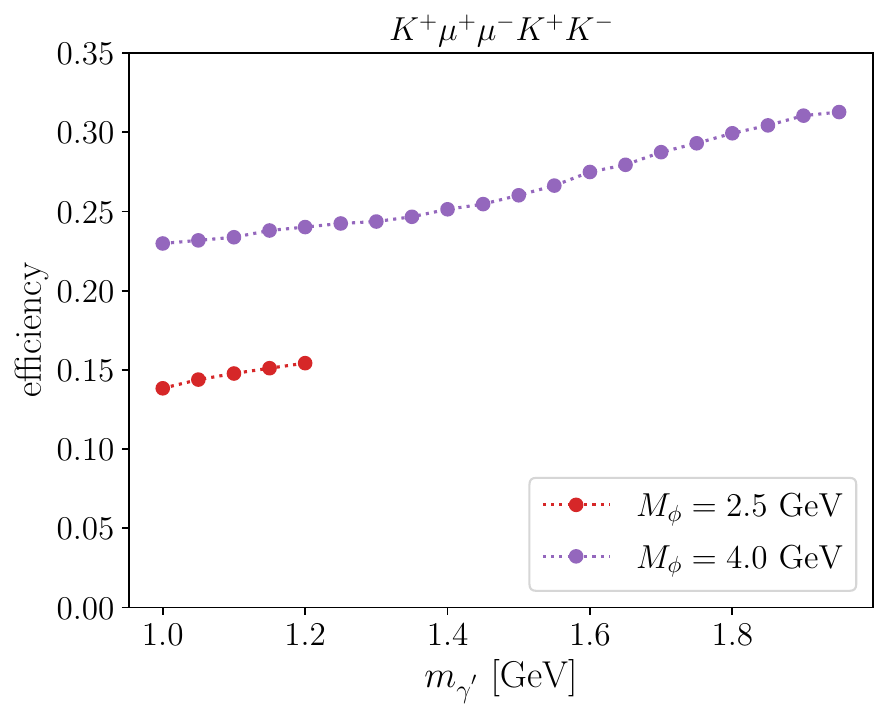}
    \includegraphics[width=0.45\linewidth]{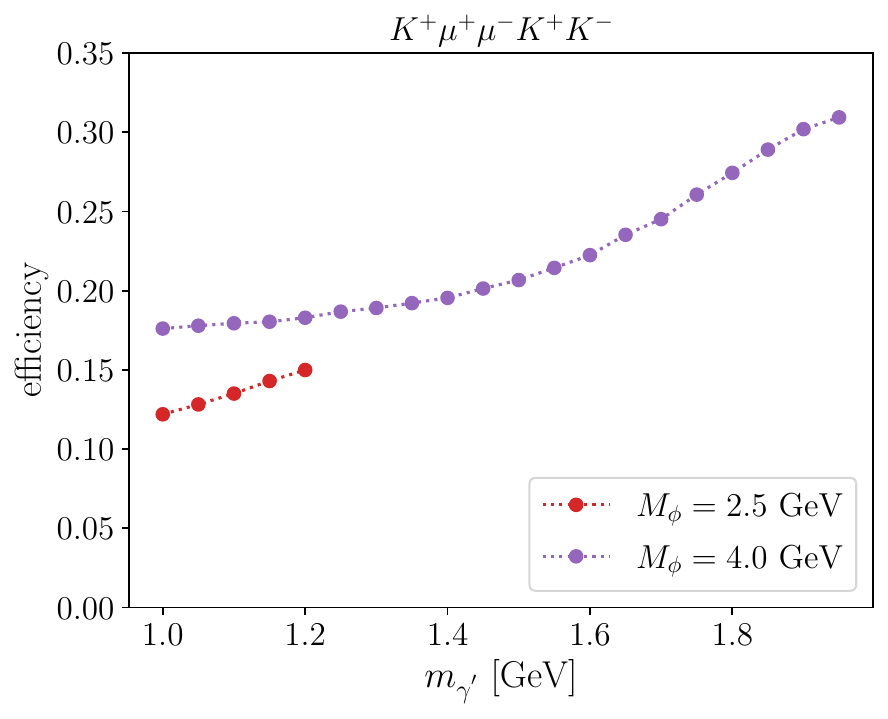}
    \caption{The same format as in  Fig.~\ref{fig:eff_figs_Kpluseeee} but for the $K^{+}\mu^{+}\mu^{-}K^{+}K^{-}$ final state.
    }
    \label{fig:eff_figs_KplusmmKK}
\end{figure}

\begin{figure}[htbp]
    \centering
    \includegraphics[width=0.45\linewidth]{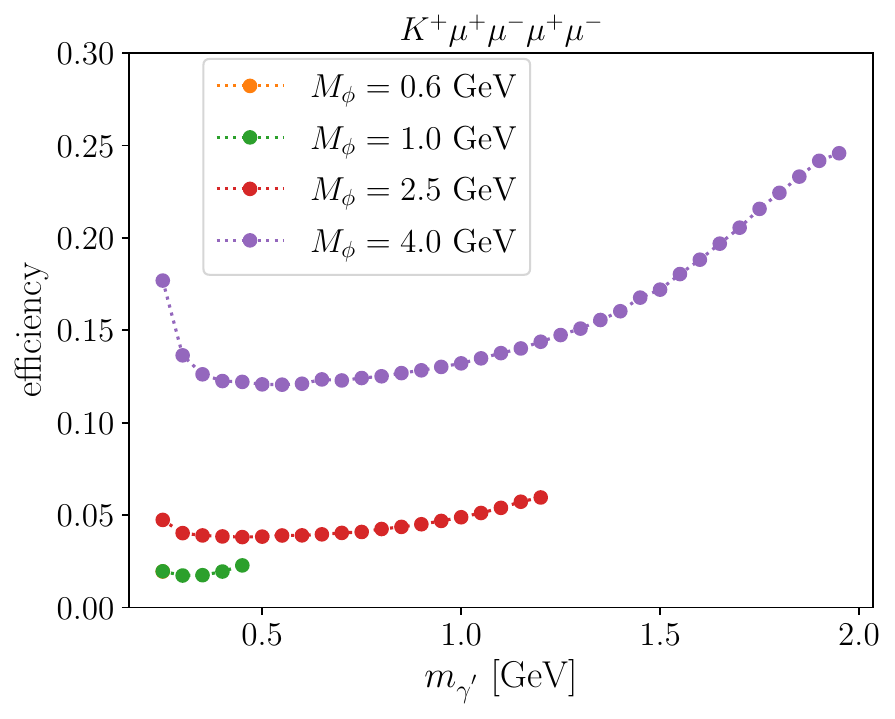}
    \includegraphics[width=0.45\linewidth]{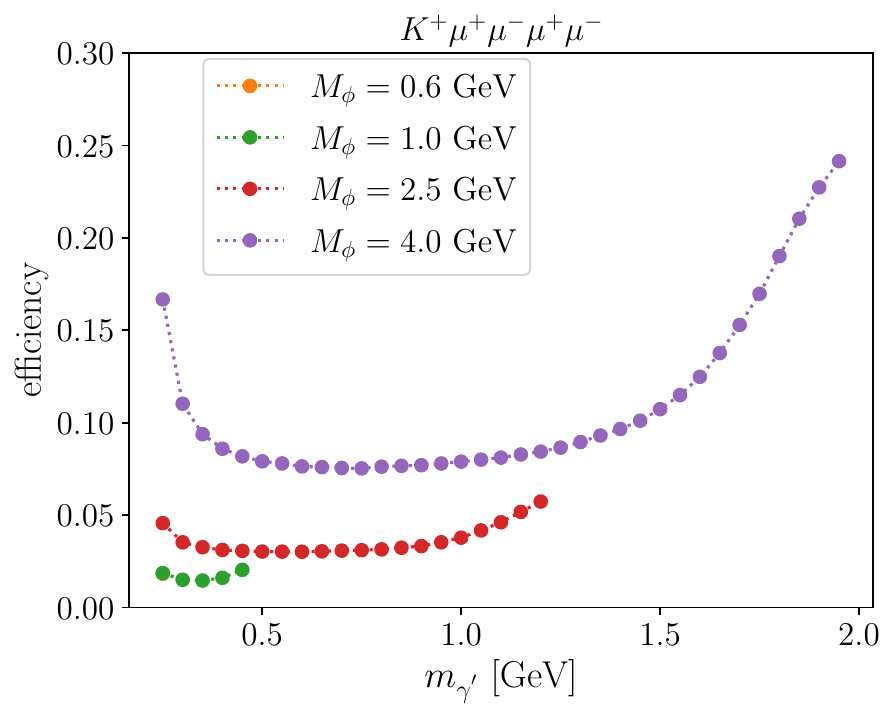}
    \caption{The same format as in  Fig.~\ref{fig:eff_figs_Kpluseeee} but for the $K^{+}\mu^{+}\mu^{-}\mu^{+}\mu^{-}$ final state.
    }
    \label{fig:eff_figs_Kplusmmmm}
\end{figure}

\begin{figure}[htbp]
    \centering
    \includegraphics[width=0.45\linewidth]{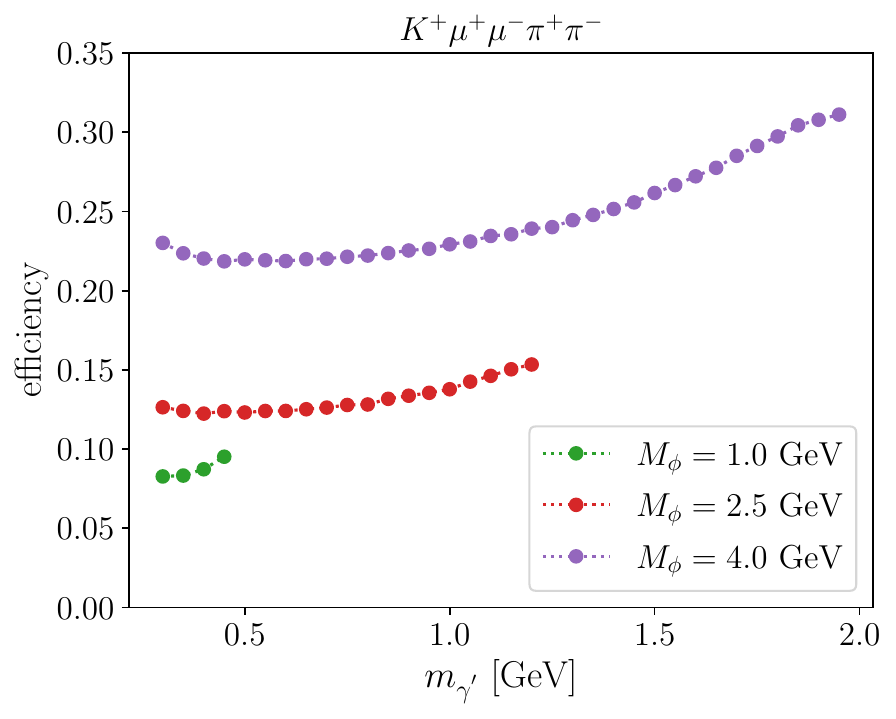}
    \includegraphics[width=0.45\linewidth]{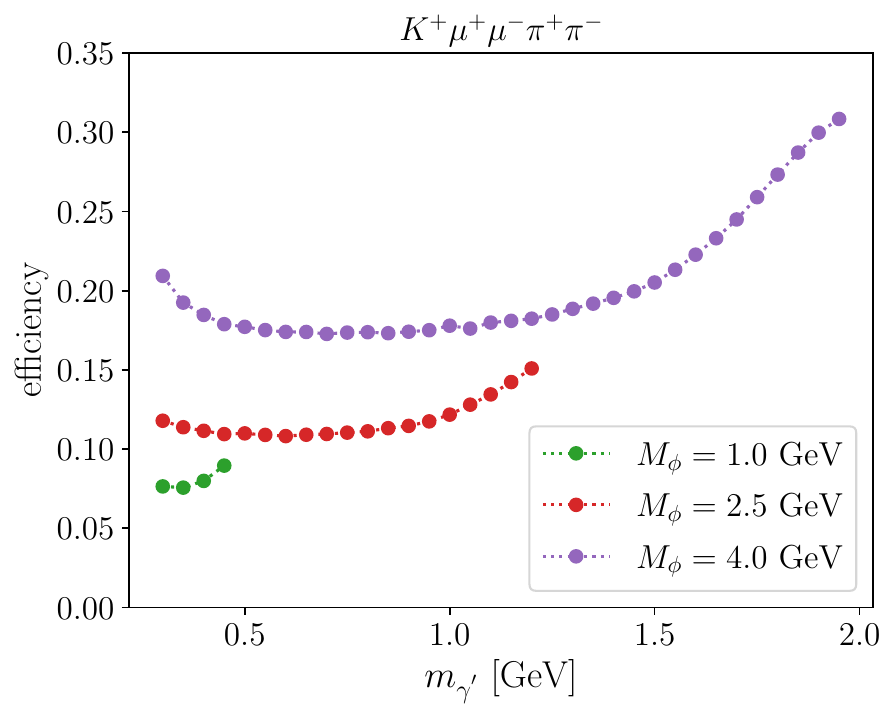}
    \caption{The same format as in  Fig.~\ref{fig:eff_figs_Kpluseeee} but for the $K^{+}\mu^{+}\mu^{-}\pi^{+}\pi^{-}$ final state.
    }
    \label{fig:eff_figs_Kplusmmpp}
\end{figure}

\begin{figure}[htbp]
    \centering
    \includegraphics[width=0.45\linewidth]{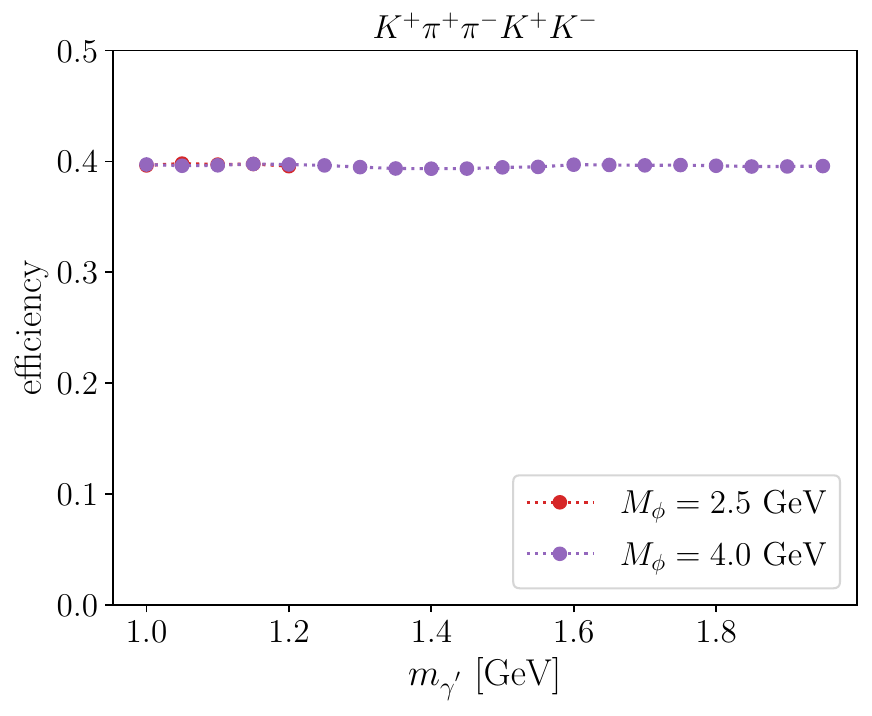}
    \includegraphics[width=0.45\linewidth]{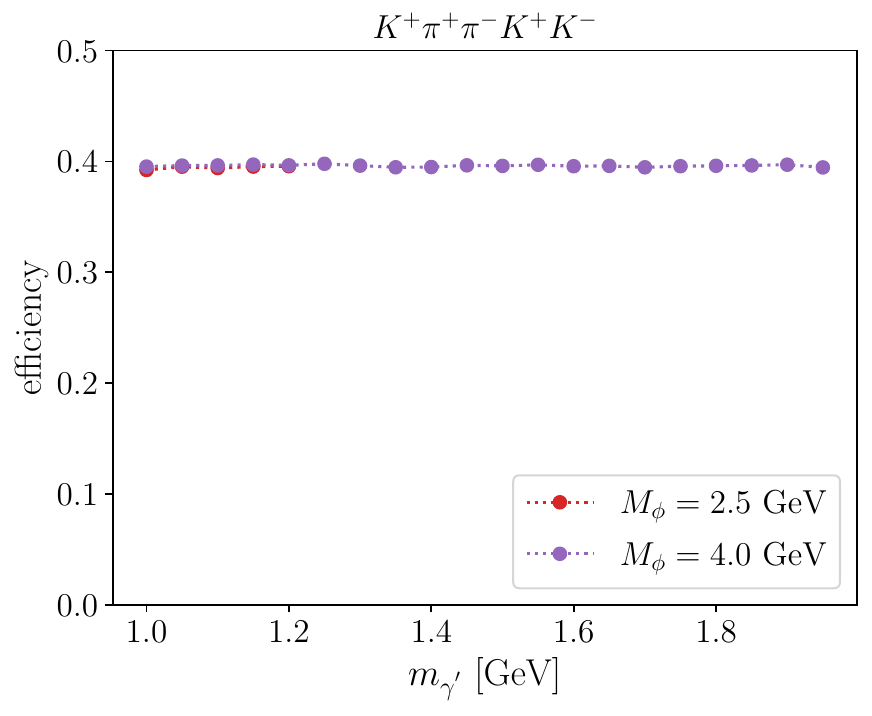}
    \caption{The same format as in  Fig.~\ref{fig:eff_figs_Kpluseeee} but for the $K^{+}\pi^{+}\pi^{-}K^{+}K^{-}$ final state.
    }
    \label{fig:eff_figs_KplusppKK}
\end{figure}

\begin{figure}[htbp]
    \centering
    \includegraphics[width=0.45\linewidth]{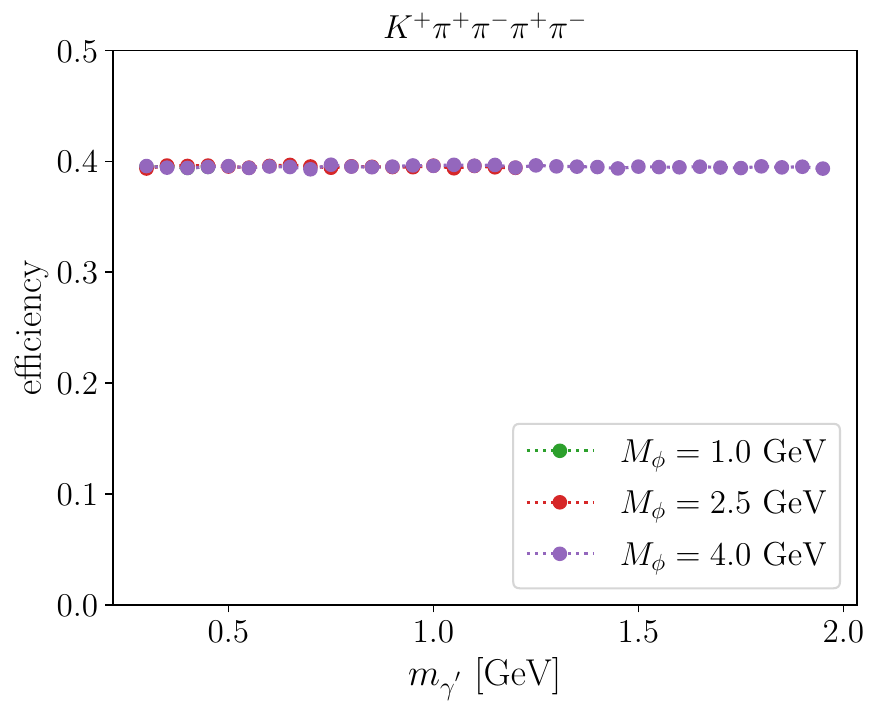}
    \includegraphics[width=0.45\linewidth]{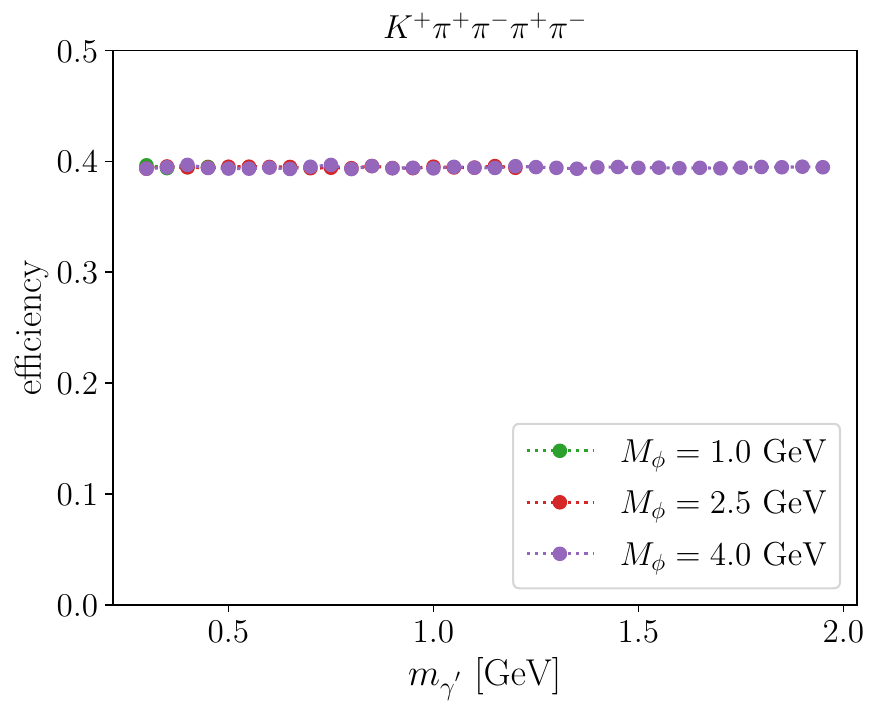}
    \caption{The same format as in  Fig.~\ref{fig:eff_figs_Kpluseeee} but for the $K^{+}\pi^{+}\pi^{-}\pi^{+}\pi^{-}$ final state.
     }
    \label{fig:eff_figs_Kpluspppp}
\end{figure}

\begin{figure}[htbp]
    \centering
    \includegraphics[width=0.45\linewidth]{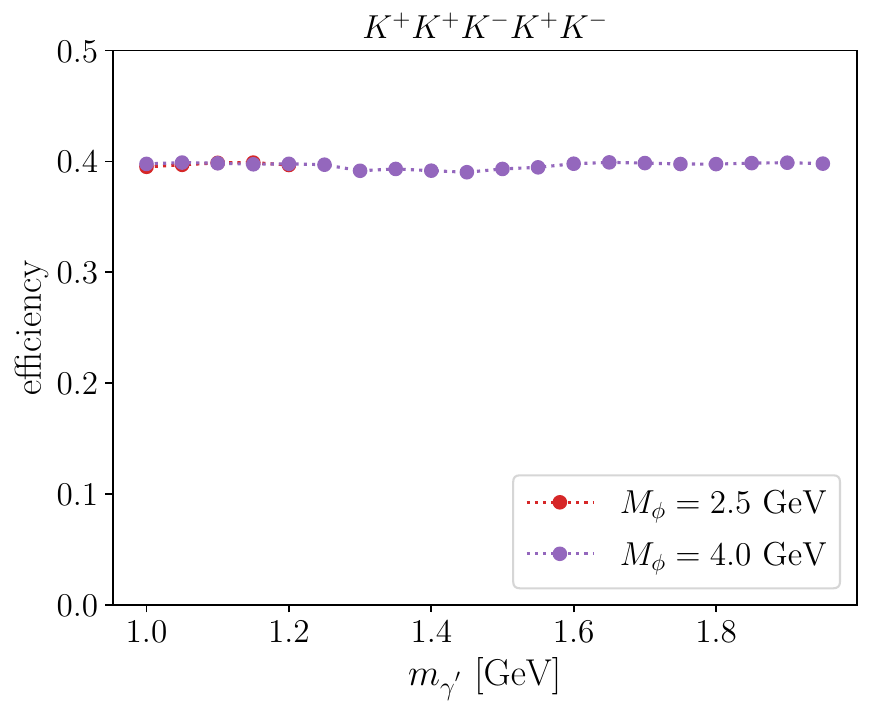}
    \includegraphics[width=0.45\linewidth]{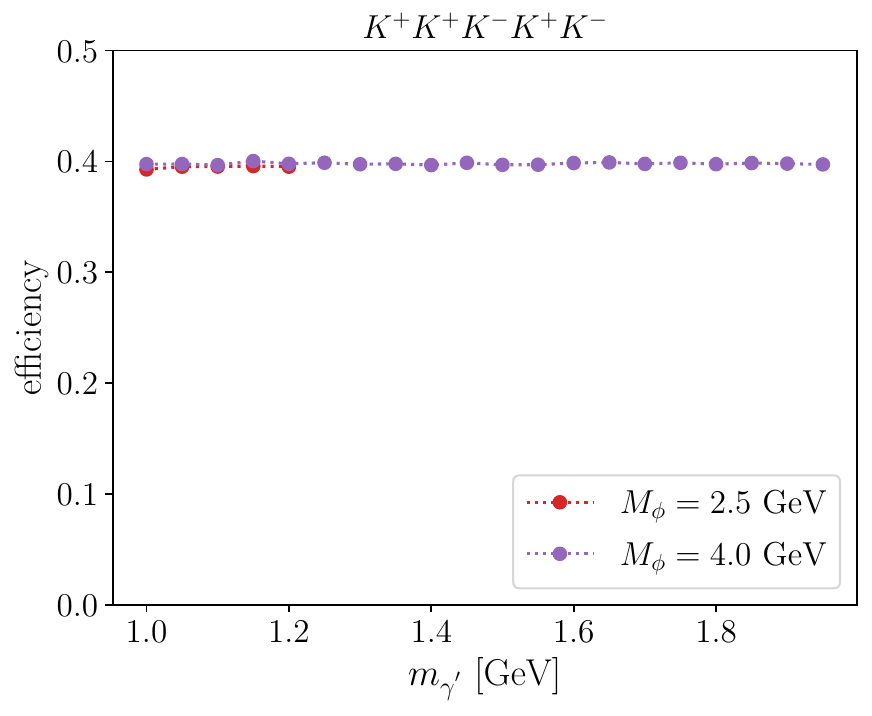}
    \caption{The same format as in Fig.~\ref{fig:eff_figs_Kpluseeee} but for the $K^{+}K^{+}K^{-}K^{+}K^{-}$ final state.}
    \label{fig:eff_figs_KplusKKKK}
\end{figure}

Table~\ref{table:eff_eid} shows the electron-identification efficiency, based on use of the calorimeter only, as a function of momentum and polar angle. 
The muon-identification efficiency is presented in Table~\ref{table:eff_muid}.
The final particle-identification efficiency $\varepsilon_{ij}^{\text{PID}}$ is shown in Figs.~\ref{fig:eff_figs_Kpluseeee}-\ref{fig:eff_figs_KplusKKKK} for each final state and for different masses of the dark Higgs and the dark photon.

 \begin{table}[htbp]
    \centering
    \begin{tabular}{c|r r r}
        $e^{\pm}$ efficiency                & $p<0.3$  & $0.3<p<1.0$  & $p>1.0$ \\
        \hline
        $\theta<17^{\circ}$                 & $0.00$   & $0.00$       & $0.00$  \\
        $17^{\circ}<\theta<31.4^{\circ}$    & $0.00$   & $0.81$       & $0.96$  \\
        $31.4^{\circ}<\theta<32.2^{\circ}$  & $0.00$   & $0.68$       & $0.81$  \\
        $32.2^{\circ}<\theta<128.7^{\circ}$ & $0.00$   & $0.81$       & $0.96$  \\
        $128.7^{\circ}<\theta<130.7^{\circ}$& $0.00$   & $0.68$       & $0.81$  \\
        $130.7^{\circ}<\theta<150^{\circ}$  & $0.00$   & $0.77$       & $0.91$  \\
        $150^{\circ}<\theta$                & $0.00$   & $0.00$       & $0.00$  \\

    \end{tabular}
    \caption{Simplified electron-identification efficiency with respect to momentum and polar angle~\cite{Belle2physbook}.}
    \label{table:eff_eid}
\end{table}

\begin{table}[htbp]
    \centering
    \begin{tabular}{c|r r r}
        $\mu^{\pm}$ efficiency          & $p<0.6$  & $0.6<p<1.5$  & $p>1.5$ \\
        \hline
        $\theta<17^{\circ}$             & $0.00$   & $0.00$       & $0.00$  \\
        $17^{\circ}<\theta<110^{\circ}$ & $0.00$   & $0.81$       & $0.96$  \\
        $110^{\circ}<\theta<130^{\circ}$& $0.00$   & $0.76$       & $0.90$  \\
        $130^{\circ}<\theta<150^{\circ}$& $0.00$   & $0.81$       & $0.96$  \\
        $150^{\circ}<\theta$            & $0.00$   & $0.00$       & $0.00$  \\
    \end{tabular}
    \caption{Simplified muon-identification efficiency with respect to momentum and polar angle~\cite{Belle2physbook}.}
    \label{table:eff_muid}
\end{table}

\bibliographystyle{JHEP}
\bibliography{bib}

\end{document}